\documentclass[%
 reprint,
superscriptaddress,
 amsmath,amssymb,
 aps,
]{revtex4-1}

\usepackage{graphicx}% Include figure files
\usepackage{dcolumn}% Align table columns on decimal point
\usepackage{bm}
\usepackage{physics}% bold math
\usepackage{hyperref}% add hypertext capabilities
\usepackage{siunitx}
\usepackage[caption=false]{subfig}

\begin{document}

\preprint{APS/123-QED}

\title{Einstein vs Hawking:\\
Black hole binaries and cosmological expansion}% Force line breaks with \\

%\author{Baptiste Blachier}
%\email{baptiste.blachier@ens-lyon.fr}
% \affiliation{Laboratoire de Physique Subatomique et de Cosmologie, Univ. Grenoble-Alpes, CNRS/IN2P3 \\ 53, avenue des Martyrs, 38026 Grenoble cedex, France}
% \affiliation{Department of Physics, \'Ecole Normale Supérieure de Lyon, 69364 Lyon, France}
%\author{Aurélien Barrau}%
% \email{barrau@lpsc.in2p3.fr}
%\author{Killian Martineau}
%\email{martineau@lpsc.in2p3.fr}
%\author{Cyril Renevey}
%\email{renevey@lpsc.in2p3.fr}
%\affiliation{Laboratoire de Physique Subatomique et de Cosmologie, Univ. Grenobletypical -Alpes, CNRS/IN2P3 \\ 53, avenue des Martyrs, 38026 Grenoble cedex, France}

\author{Aurélien Barrau}%
 \email{barrau@lpsc.in2p3.fr}
\affiliation{Laboratoire de Physique Subatomique et de Cosmologie, Univ. Grenoble-Alpes, CNRS/IN2P3 \\ 53, avenue des Martyrs, 38026 Grenoble cedex, France}

\author{Baptiste Blachier}
\email{baptiste.blachier@ens-lyon.fr}
 \affiliation{Laboratoire de Physique Subatomique et de Cosmologie, Univ. Grenoble-Alpes, CNRS/IN2P3 \\ 53, avenue des Martyrs, 38026 Grenoble cedex, France}
 \affiliation{Department of Physics, \'Ecole Normale Supérieure de Lyon, 69364 Lyon, France}
\author{Maxime Lahlou}
\email{mlahlou@princeton.edu}
 \affiliation{Princeton University, Princeton, NJ, 08544, USA}

\author{Andrew Liu}
\email{andrew.s.liu@princeton.edu}
 \affiliation{Laboratoire de Physique Subatomique et de Cosmologie, Univ. Grenoble-Alpes, CNRS/IN2P3 \\ 53, avenue des Martyrs, 38026 Grenoble cedex, France}
 \affiliation{Princeton University, Princeton, NJ, 08544, USA}
 
\author{Killian Martineau}
\email{martineau@lpsc.in2p3.fr}
\affiliation{Laboratoire de Physique Subatomique et de Cosmologie, Univ. Grenoble-Alpes, CNRS/IN2P3 \\ 53, avenue des Martyrs, 38026 Grenoble cedex, France}

\date{\today}% It is always \today, today,

\begin{abstract}
This note aims at investigating two different situations where the classical general relativistic dynamics competes with the evolution driven by Hawking evaporation. We focus, in particular, on binary systems of black holes emitting gravitational waves and gravitons, and on the cosmological evolution when black holes are immersed in their own radiation bath. Several non-trivial features are underlined in both cases.
\end{abstract}

\maketitle

\section{\label{sec:intro}Introduction}

An uncharged and non-rotating black hole (BH) in vacuum is fully described by the Schwarzschild solution to the Einstein's field equations. According to Birkhoff's theorem, this solution is actually unique \cite{birk}. The main expected standard physics addition to this general.relativistic picture is Hawking evaporation \cite{Hawking:1975vcx}, which leads to an explosive mass loss when the BH is light enough. Neglecting higher-order effects, the entire geometric picture is therefore encoded in the treatment provided by general relativity supplemented by the Hawking process. This is obviously well-known \cite{Jacobson:2003vx}. However, we shall show in this work that quite unexpected behaviors do emerge in some circumstances, in particular when both effects are simultaneously at play.\\

In this article, we intend to focus on two specific situations where the evolution driven by classical effects competes with the one induced by the Hawking mechanism. The aim of this study is not to study new physics effects, nor to provide a framework for explicit phenomenological investigations. We simply try to clarify, at the theoretical level, the subtle interplay between classical ({\it i.e.} described by Einstein gravity) and semi-classical ({\it i.e.} accounted for by the Hawking mechanism) effects. As we will demonstrate in the following, many complicated situations can be encountered. Our objective is to put the emphasis on cases where the dominant contribution could not have been easily guessed and to provide a detailed quantitative analysis of the subtleties entering the game. Although assuming that the considered system is isolated is obviously a strong hypothesis, we insist that within this restriction our analysis is complete. The relative amplitude of the considered effects depends upon the initial values of the masses and on the orbital separation. We provide an exhaustive investigation considering all cases.\\

The first case we consider consists of a binary system of two BHs. Energy is inevitably radiated away by gravitational waves (although unquestionably correct let us underline that this apparently simple statement is actually {\it not} simple \cite{Gomes:2023xda}). This makes the system inspiral. However, simultaneously, the Hawking radiation induces a mass decrease and this makes the system outspiral. Among other particles, spin-2 bosons -- referred to as gravitons in the following -- are emitted. We investigate how spacetime excitations produced by gravitational waves compare with those triggered by gravitons. In particular, we show that the total energy radiated in the form of gravitons is always greater than the one carried away by gravitational waves. We also compare the associated frequencies and derive general results. We take this opportunity to provide some clarifications about the highly non-trivial trajectories.\\

The second situation we consider corresponds to the cosmological expansion if the Universe is filled with a gas of black holes. If space is maximally symmetric, the full general relativistic geometrical picture is encoded in Friedmann's equation for a matter fluid. However, if the BHs evaporate, things get more involved. On the one hand, the cosmic expansion favors matter over radiation (as the dilution is faster for radiation). But, on the other hand, matter gets converted -- at an increasing rate -- into radiation by the Hawking process. Who wins and why? We study in details the situation and provide new analytical results.\\

Throughout all this work, we put the emphasis on clarity. We also keep the full SI units, making numerical estimates, if required, easier than in Planck units. 

\section{\label{sec:sec2}Binary system of evaporating black holes}

\subsection{The general picture}

In principle, the dynamics of an isolated binary system of Schwarzschild black holes is fully determined by both the emission of gravitational waves and the mass loss induced by the evaporation. Depending on the mass and on the distance between the BHs, one of these effects can, of course, be very strongly dominant over the other. Considering {\it a priori} both phenomena is however the rigorous approach to get an exhaustive picture of what happens in vacuum. In addition, we shall precisely show in the following that both phenomena can simultaneously play an important role. The relevant mass range will be precisely calculated. From a phenomenological perspective, only primordial black holes -- for which accretion is anyway negligible -- are relevant in this case. We, however, keep our study at the general theoretical level.
As gravitational waves tend to induce an inspiralling trajectory whereas the mass decrease makes the system outspiral, this obviously triggers a competition between both effects. This has been investigated recently in \cite{Blachier:2023ygh}, assuming circular orbits and performing the calculations at the lowest meaningful order. The same hypotheses will be made in this work.\\

Let us make a few comments about the circularity hypothesis. First, it is well known that  back-reaction induces a circularization of the orbit. The eccentricity decreases quite fast and, for all realistic situations, becomes negligible long before the coalescence phase \cite{Maggiore:2007ulw}. 
Second, the third time derivatives of the second mass moments $M_{ij} \equiv \mu x^{i}(t) x^{j}(t)$ can be computed for elliptic orbits \cite{Holgado:2019ndl}, including the additional contributions arising from the time-varying masses along with those produced by the orbital motion. These additional terms involve an angular dependency upon the true anomaly $\psi$ and the eccentricity $e$ of the elliptic orbit but these contributions are all bounded to numerically evolve between 0 and 1. What is interesting is that, at the dimensional level, when the two masses of the binary system are equal (let us denote them by $m$), these terms scale as $\dot{m} \Omega_{0}^{2}a^{2}$, $\ddot{m} \Omega_{0}a^{2}$, and $\dddot{m} a^{2}$, with $a$ the semi-major axis of the ellipse and $\Omega_{0}^{2} \equiv 2Gm/a^{3}$  the fundamental frequency of the emitted gravitational waves for elliptic orbits. For a generic value of the eccentricity $e$, all harmonics contribute to the frequency spectrum, \textit{i.e.} there are waves radiated at all frequencies $\Omega_{l} = l \Omega_{0}$ for all integer values of $l$. \\

On the other hand, the dimensional prefactor appearing in the classic formula of Peter and Mathews \cite{Peters:1963ux} when the masses are taken constant (which is obviously also present in the formulae that generalize this situation to time-varying masses) is $m \Omega_{0}^{3} a^{2}$. Consequently, as long as
\begin{equation}
    \forall \quad 0 < n \leq 3, \qquad \vert m^{(n)}(t)\vert \ll m \Omega_{0}^{n},
\end{equation}
with $n$ an integer, the corrective terms can be neglected (see appendix). In fact, this condition is the exact analog of the one derived in \cite{Blachier:2023ygh}, the so-called \textit{slowly varying mass condition}, which imposes that  the mass of the bodies is slowly varying when compared with the typical orbital evolution $\Omega_{0}$ of the elliptic binary system. Then, over the temporal window on which the average is performed -- that is to say on a few characteristic periods of the gravitational waves -- the mass remains nearly constant and equal to $m(t)$. All in all, this means that if this hypothesis holds during the Hawing process (which is the case except during the dramatic end), if one were wishing to take into account the non-circularity at the beginning of the process, the power radiated should simply be corrected by the classic factor \cite{Peters:1963ux}
\begin{equation}
    f(e)=\frac{1}{(1-e^2)^{\frac{7}{2}}}\left(
1+\frac{73}{24}e^2+\frac{37}{96}e^4\right).
\end{equation}
This, however, does not change the order of magnitude of our conclusions.\\

The two-body problem with variable mass was initially considered in \cite{HADJIDEMETRIOU196634, Verhulst1975}. Using the conservation of the angular momentum, it is straightforward to show that the equation of motion reads
\begin{equation}
%    \frac{\dot{R}}{R} = - 3\frac{\dot{m}}{m} ,
m\dot{R}=-3\dot{m}R,
    \label{eq:dotR}
\end{equation}
where $R$ is the orbital separation and $m$ is the (variable) mass of each black hole.  We assume that the BHs  have the same mass as this is the most formally interesting case and as several models naturally lead to nearly monochromatic mass spectra (see \cite{Carr:2020gox} for a review of formation mechanisms of primordial black holes). This work is mostly conceptual but if it were to have any phenomenological relevance, the focus should clearly be on primordial black holes (PBHs) as the mass loss rate would otherwise be too small to change substantially the trajectory. Hence, the reference to a mass spectrum, as PBHs are expected to be formed in the early universe by phenomena possibly involving a continuum of scales. Should the BHs making the binary system present a high mass hierarchy, the factor 3 entering Eq. \eqref{eq:dotR} would have to be replaced by a factor 2 or 1 (depending on which BH is actually loosing mass). 

On the other hand, the equation of motion for a binary system (with fixed masses) emitting gravitational waves reads \cite{Maggiore:2007ulw}
\begin{equation}
    \dot{R} R^3= -\frac{128}{5} \frac{G^{3}}{c^{5}} m^{3},
\label{eq:ED_general}
\end{equation}
where $c$ is the speed of light, and $G$ is the gravitational constant. Both equations easily combine \cite{Enander_2010}, leading to:

\begin{equation}
    \dot{R} = -\frac{128}{5} \frac{G^{3}}{c^{5}} \frac{m^{3}}{R^{3}} - 3\frac{\dot{m}}{m}R,
\label{eq:ED_general}
\end{equation}
which is the general differential equation relevant for our problem. 

In \cite{Blachier:2023ygh}, we have shown that this quite simple equation leads to a surprisingly rich landscape of behaviors. \\

We now focus on a specific question: which process releases the highest amount of power in the form of spacetime vibrations? Otherwise stated: how does the gravitational wave power, emitted by the orbiting black holes, compare with the graviton power, due to the Hawking evaporation? The answer is not {\it a priori} obvious and deserves a specific study. The Hawking process converts directly mass into radiation which represents a kind of ultimate efficiency. However, when the system merges the emission of gravitational waves can also extract a huge amount of energy from the orbit. Who wins? Why and when? Incidentally, how does the frequency of gravitons compare with the one of gravitational waves?

A fundamental remark is in order at this stage. Equation \eqref{eq:ED_general} is deeply asymmetric as the evaporation process is unaffected by the emission of gravitational waves whereas the orbit is drastically influenced by the mass loss induced by the Hawking effect. This makes the situation quite tricky.\\

Basically, Hawking evaporation is characterized by a (nearly) thermal spectrum with temperature \cite{Hawking:1975vcx}
\begin{equation}
	T = \frac{\hbar c^{3}}{8\pi G k_{\mathrm{B}} m},
\end{equation}
where $\hbar$ is the Planck constant, and $k_{\mathrm{B}}$ is the Boltzmann constant. As the BH evaporates it gets lighter and lighter, hence hotter and hotter: the process is explosive and characterized by a negative heat capacity. Importantly, being gravitational, the evaporation is nearly democratic in the sense that the BH couples equally to all the fundamental degrees of freedom. In principle, the spin does play a role both in the accurate expression of the spectrum and in the so-called greybody factors which account for slight modulations with respect to the standard blackbody behaviour \cite{Page:1976df}. Those effects however remain weak and will be neglected in this study. In addition, we assume that all particles are emitted at the same energy $E=\zeta k_BT$. This approximates the thermal distribution by its maximum. For integer spins, $\zeta \approx 1.6$ \cite{MacGibbon_PBH}. As gravitons represent 2 degrees of freedom among the 128 of the Standard Model (SM), they will constitute $\xi=1/64$ of the total emitted flux. This assumes that the temperature of the black hole is higher than all the SM masses. If the BH is large enough so that this is not true anymore, this assumption will slightly underestimate the relative importance of gravitons. This however does not change any conclusion.\\

The dynamics of the evaporation can easily be obtained by integrating the instantaneous Hawking spectrum. This leads to:
\begin{equation}
    \dot{m} = - \frac{\alpha_{\mathrm{H}}}{m^{2}},
\end{equation}
where $\alpha_{\mathrm{H}}$ accounts for the number of particles and their spins (see, {\it i.e.}, \cite{Halzen:1991uw} for numerical values). Consistently with the previously mentioned hypotheses, we take $\alpha_{\mathrm{H}}$ to be constant, which is a good approximation.  This integrates into 
\begin{equation}
    m(t) = m_{0} \left(1 - \frac{t}{t_{\mathrm{ev}}}\right)^{1/3},
\label{eq:hawking_mass}
\end{equation}
where $m_{0}$ is the initial mass and $t_{\mathrm{ev}} \equiv m_{0}^{3}/(3\alpha_{\mathrm{H}})$ corresponds to the time it takes to the BH to fully evaporate.\\

The dynamics of the orbit is obviously more involved as it depends both on the emission of gravitational waves {\it and} on the mass loss. We have shown in \cite{Blachier:2023ygh} that for any initial mass  $m_{0}$, it exists three different regimes, depending on the initial orbital separation $R_{0}$. Let us define
\begin{equation}
    R_{1} \equiv \left(\frac{256}{45} \frac{G^{3}}{c^{5} \alpha_{\mathrm{H}}}\right)^{1/4} m_{0}^{3/2},
\end{equation}
and
\begin{equation}
    R_{2} \equiv \left(\frac{128}{15} \frac{G^{3}}{c^{5} \alpha_{\mathrm{H}}}\right)^{1/4} m_{0}^{3/2}.
\end{equation}
When $R_{0} < R_{1}$, the system is inspiralling: the emission of gravitational waves dominates the dynamics. When $R_0>R_2$, the other way around, the system is outspiralling as the mass decrease plays the dominant role in the orbital evolution. For $R_1<R_0<R_2$, the dynamics is non-monotonic: the system first inspirals and, then, outspirals. \\

With $\alpha_{\mathrm{H}} = \SI{7.8e17}{kg^{3}.s^{-1}}$, one can numerically express $R_{1}$ and $R_{2}$ under the form 
\begin{equation}
    R_{1} = \SI{2.8}{Mpc} \left(\frac{m_{0}}{M_{\odot}}\right)^{3/2} = \SI{3.1e-5}{m} \left(\frac{m_{0}}{M_{*}}\right)^{3/2}
\end{equation}
and
\begin{equation}
    R_{2} = \SI{3.1}{Mpc} \left(\frac{m_{0}}{M_{\odot}}\right)^{3/2} = \SI{3.4e-5}{m} \left(\frac{m_{0}}{M_{*}}\right)^{3/2}
\end{equation}
where $M_{*} \equiv \SI{e12}{kg}$ is the typical mass of black holes whose time of evaporation is equal to Hubble time.

In the following, we consider the evolution from the initial time to the evaporation time if the system outspirals. We instead consider the coalescence -- or innermost stable circular orbit (ISCO), which is nearly the same -- time as the final point if the system inspirals. At the very end of the section, we extend the analysis.

\subsection{Frequencies}

Intuition is easier when considering wavelengths. For an evaporating BH, the Schwarzschild radius $R_{\mathrm{S}}=2Gm/c^2$ is the only length scale of the problem. The wavelength of the emitted gravitons should therefore be of the same order of magnitude: $\lambda_{\mathrm{gr}}\sim R_{\mathrm{S}}$. 

On the other hand, for gravitational waves, another scale enters the game: the orbital separation. Kepler's third law reads, for the orbital frequency, $\omega^{2} = 2Gm/R^{3}$, which leads to an associated gravitational wavelength $\lambda_{\mathrm{GW}}\sim \sqrt {R^3/R_{\mathrm{S}}}$ at all times.

As the orbital separation is always greater than the Schwarzschild radius (otherwise the system would have merged) it is obvious that $\lambda_{\mathrm{gr}}<\lambda_{\mathrm{GW}}$.

The order of magnitude of the ratio is $\lambda_{\mathrm{GW}}/\lambda_{\mathrm{gr}}\sim (R/R_{\mathrm{S}})^{3/2}$ and its evolution is therefore, as expected, only governed by the relative evolution of the two scales of the problem.\\

Let us now be more quantitative. It is possible to calculate the exact analytical expression for both frequencies as a function of time. Relying on the $R(t)$ expression derived in \cite{Blachier:2023ygh}, one can easily obtain for gravitational waves:
\begin{equation}
    \omega_{\mathrm{GW}} = 2\omega_{0}\left(\frac{t}{t_{\mathrm{ev}} - t}\right)^{\frac{5}{3}}\left(1 + \frac{1}{6}\frac{t_{\mathrm{ev}}}{t_{\mathrm{cc}}}\left[\left(1 - \frac{t}{t_{\mathrm{ev}}}\right)^6-1\right]\right)^{-\frac{3}{8}},
    \label{eq:freq_gw}
\end{equation}
where $\omega_{0} \equiv 2Gm_{0}/R_{0}^3$ is the initial orbital frequency of the system and \begin{equation}
    t_{\mathrm{cc}} \equiv \frac{5}{512}\frac{c^5R_{0}^4}{G^3m_{0}^3}
\end{equation}
corresponds to the time of coalescence if the masses were constant (which is therefore always smaller than the actual value). 

On the other hand, the frequency for gravitons is simply given by:
\begin{equation}
    \omega_{\mathrm{gr}} = \frac{\zeta c^3}{8\pi Gm_{0}(1 - \frac{t}{t_{\mathrm{ev}}})^{\frac{1}{3}}}.
    \label{eq:freq_Gravitons}
\end{equation}

When the system outspirals (which is always the case when $R_0>R_2$ and which happens soon after the initial time for $R_1<R_0<R_2$), it is clear that the frequency given by Eq. \eqref{eq:freq_gw} is always smaller than the one of Eq. \eqref{eq:freq_Gravitons}. As the BHs drift apart, the ratio $\omega_{\mathrm{gr}}/\omega_{\mathrm{GW}}$ indeed increases (starting with an initial value greater than 1) as the graviton frequency only increases with time whereas the gravitational wave frequency always decreases with time. When the system approaches $t_{\mathrm{ev}}$, that is when the mass of the BHs approaches zero, the graviton frequency tends to infinity -- or, at least, to the Planck value -- whereas the gravitational wave frequency vanishes. \\

However, when the system is inspiralling (which is always the case when $R_0<R_1$ and which holds briefly after the initial time for $R_1<R_0<R_2$), both frequencies increase with time. This is the non-trivial case. This time, the ratio  $\omega_{\mathrm{gr}}/\omega_{\mathrm{GW}}$ does decrease (obviously still starting from a value greater than one).

\begin{figure}
    \centering
    \includegraphics[width=0.45\textwidth]{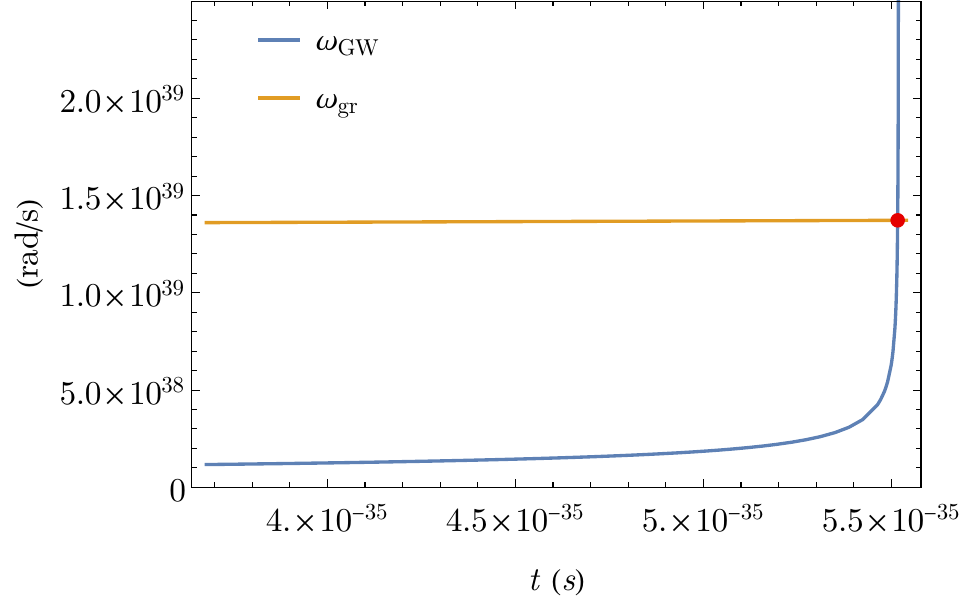}
    \caption{Typical evolution of the frequencies close to the merging. It can be seen that the gravitational wave frequency $\omega_{\mathrm{GW}}$ varies much faster than the graviton frequency $\omega_{\mathrm{gr}}$. The curve is extended after the red dot corresponding to the intersection but is not physical anymore.}
    \label{fig:intersect_freq_merging}
\end{figure}

The naive dimensional analysis presented at the beginning of this section suggests that both frequencies converge when the system coalesces. As the orbital frequency cannot even be defined beyond the ISCO, we evaluate the final quantities at this point. In practice, going from $R=12Gm/c^2$, that is the ISCO, to $R=4Gm/c^2$ does not change the orders of magnitude.
At ISCO, we obtain:
\begin{equation}
 \omega_{\mathrm{GW}}(t_{\mathrm{ISCO}}) = \frac{1}{6\sqrt{6}}\frac{c^3}{Gm_{\mathrm{f}}} \qq{and} \omega_{\mathrm{gr}}(t_{\mathrm{ISCO}}) = \frac{\zeta c^3}{8\pi Gm_{\mathrm{f}}},
 \label{eq:ISCO_frequencies}
\end{equation}
where $m_{\mathrm{f}} \equiv m(t_{\mathrm{ISCO}})$.
The ratio is
\begin{equation}
    \frac{\omega_{\mathrm{gr}}}{\omega_{\mathrm{GW}}}(t_{\mathrm{ISCO}}) = \left(\frac{27\zeta^2}{8\pi^2}\right)^{1/2} = \mathcal{O}(1).
\end{equation}
It is, as expected, of order unity and does not depend on the mass. This means that both frequencies converge at the ISCO (or merging) with a wavelength of the order of the only scale of the problem which is then the Schwarzschild radius. As shown in Fig. \ref{fig:intersect_freq_merging}, the gravitational wave frequency, which starts from a much lower value, generically increases much faster than the graviton frequency.\\

To summarize, the frequency of gravitons is always higher than the one of gravitational waves. The ratio $\omega_{\mathrm{gr}}/\omega_{\mathrm{GW}}$ decreases if the orbital separation decreases and both frequencies become roughly equal at the ISCO. The other way around, the ratio diverges at finite time if the system outspirals. This is not surprising, although the details might not have been guessed so easily.

\subsection{Power}

\subsubsection{Gravitational waves}

A more subtle question is the one of the comparison between the power radiated by the system as gravitational waves and as gravitons. Let us first focus on the gravitational wave power alone and ask a preliminary question: how does it vary with time? In the approximation scheme previously mentioned, the energy radiated by unit time reads \cite{Maggiore:2007ulw}
\begin{equation}
    P_{\mathrm{GW}} = \frac{64}{5} \frac{G^{4}}{c^{5}} \frac{m^{5}}{R^{5}},
\label{eq:Pgw}
\end{equation}
where both $R$ and $m$ are time-dependent (the evolution of the first being, in addition, affected by the second).

\begin{figure}
    \centering
    \includegraphics[width=0.45\textwidth]{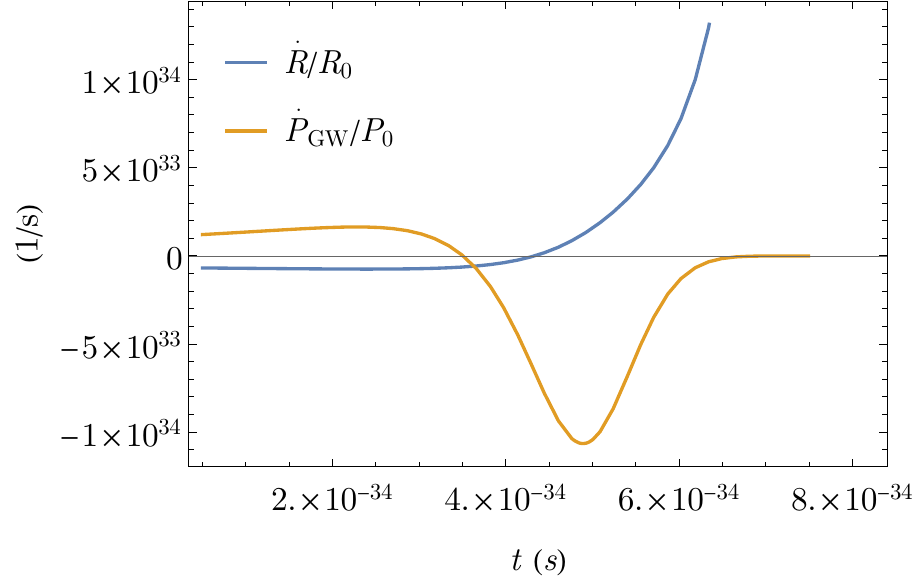}
    \caption{Normalized derivative of the orbital separation and of the power radiated as gravitational waves. It can be seen that they can be both negative at the same time, which corresponds to a situation where the power released as gravitational wave decreases whereas the system inspirals.}
    \label{fig:Inspiraling_Power_Decrease}
\end{figure}

When the system is outspiralling, it is obvious that the gravitational wave power decreases. However, when the system inspirals, the situation gets tricky: the decrease of $R$ clearly tends to increase the power but the decrease of $m$ plays in the opposite direction. Gathering all terms involved in the power, one is led to:

\begin{equation}
      P_{\mathrm{GW}} = P_{0}\left(1 - \frac{t}{t_{\mathrm{ev}}}\right)^{\frac{20}{3}}\left(1 + \frac{1}{6}\frac{t_{\mathrm{ev}}}{t_{\mathrm{cc}}}\left[\left(1 - \frac{t}{t_{\mathrm{ev}}}\right)^6-1\right]\right)^{-\frac{5}{4}}
      \label{eq:Power_gw}
\end{equation}
with $P_{0} = 64G^4m_{0}^5/(5c^5R_{0}^5).$ Figure \ref{fig:P/P_{0}vt(s)} shows the typical evolution of $ P_{\mathrm{GW}}$ and exhibits three main regimes, similarly to the trends that were presented in Ref. \cite{Blachier:2023ygh} for the orbital separation and frequency.
The condition $\dot P > 0$ is equivalent to
\begin{equation}
    \frac{\dot m}{m} > \frac{\dot R}{R}.
    \label{eq:inequal_Pdot}
\end{equation}
This translates into $R<R_{\mathrm{c}}$ with
\begin{equation}
    R_{\mathrm{c}} \equiv \left(\frac{32G^3}{5\alpha_{\mathrm{H}}c^5}\right)^\frac{1}{4}m^{3/2}. 
\end{equation}
This can be numerically expressed as
\begin{equation}
    R_{\mathrm{c}} = \SI{2.9}{Mpc} \left(\frac{m_{0}}{M_{\odot}}\right)^{3/2} = \SI{3.2e-5}{m} \left(\frac{m_{0}}{M_{*}}\right)^{3/2}.
\end{equation}
Interestingly, this lies between $R_1$ and $R_2$. It means that when $R_0>R_2$ the system is only outspiralling and $P_{\mathrm{GW}}$ is always monotonically decreasing. When $R_0<R_1$ the system inspirals and the power monotonically increases. However when $R_c<R_0<R_2$, which is a small but non-vanishing interval, the gravitational wave power does decrease at the beginning of the trajectory while the orbital separation also decreases. There is no one-to-one correspondence between the sign of $\dot P$ and the sign of $\dot{R}$ (see Fig. \ref{fig:Inspiraling_Power_Decrease} as an illustration).

Basically, this -- perhaps surprising -- behavior can be understood as follows. For most of the parameter space, the evolution of the gravitational wave power is determined by the orbital evolution. This is simply due to the fact that if the graviton power was much higher than the gravitational wave power, the system would anyway be outspiralling. It is  not possible for the binary system to inspiral while having $P_{\mathrm{gr}}\gg P_{\mathrm{GW}}$. However, the power is not what directly dictates the dynamics and this is why, in a narrow window, the power evolution might not follow the evolution of the orbit.\\

At the ISCO and merging the gravitational wave power reads
\begin{equation}
    P_{\mathrm{GW}}^{\mathrm{ISCO}}=\frac{1}{19440}\frac{c^5}{G}
    \qq{and} 
    P_{\mathrm{GW}}^{\mathrm{merger}}=\frac{1}{80}\frac{c^5}{G}.
\label{eq:finalP}
\end{equation}

This raises two important comments. First, those values do not depend on the mass. The power released at the final stage of a merger of black holes is independent of their masses: lighter BHs come closer to one another and this ``compensates" for the smaller source term. Of course, the derivative of the power depends on the mass (as $1/m$) and so does the full radiated energy. Secondly, the power value at merging is huge, reaching a nearly Planckian value.\\

%Figure \ref{fig:P/P_{0}vt(s)} shows the typical evolution of the gravitational wave power for the 3 main regimes whereas Fig. \ref{fig:Inspiraling_Power_Decrease} illustrates the fact that $P_{\mathrm{GW}}$ can increase although the system outspirals

\begin{figure}[t!]
    \centering
    \includegraphics[width=0.4\textwidth]{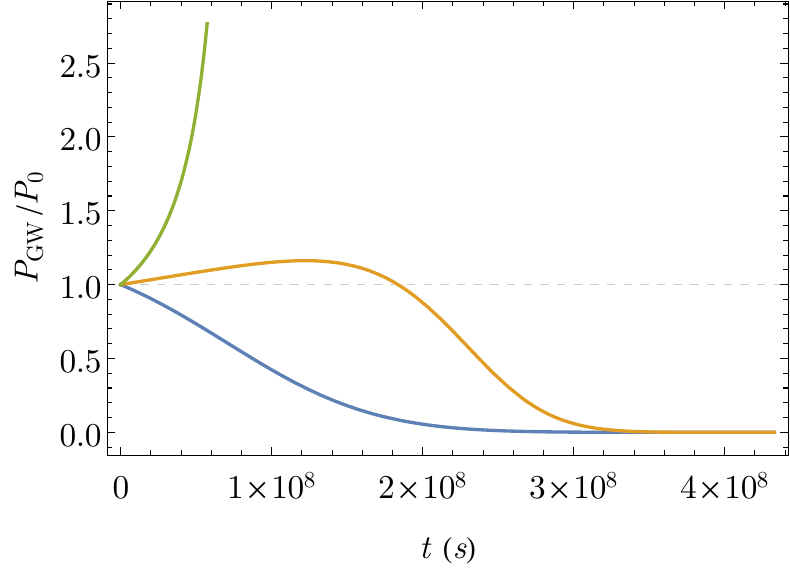}
    \caption{Typical time evolution of the power radiated as gravitational waves for three different regimes (in green, the purely inspiralling case, in blue the purely outspiralling case, and in brown an intermediate case).}
    \label{fig:P/P_{0}vt(s)}
\end{figure}

\subsubsection{Comparison between gravitational waves and gravitons}

While the time evolution of the gravitational wave power is given by Eq. \eqref{eq:Power_gw}, the graviton power reads:
\begin{equation}
P_{\mathrm{gr}} = \frac{\xi\alpha_{\mathrm{H}}c^2}{m_{0}^2(1 - \frac{t}{t_{\mathrm{ev}}})^{\frac{2}{3}}}.
\label{eq:P_G}
\end{equation}
In principle, all the needed material is therefore at disposal. 
The situation is however quite involved and the best way to get a clear idea of the main trends is to focus on the initial and final behaviors of the relevant powers.\\

Let us first mention that the ultimate fate of the system just depends upon the relative values of $R_0$ and $R_1$. The relative position with respect to $R_2$ only changes the initial behavior. 

If $R_0<R_1$, the final power in gravitational waves is given by Eq. \eqref{eq:finalP} while the final power in gravitons is 
\begin{equation}
P_{\mathrm{gr}}(t_{\mathrm{ISCO}}) = \frac{\xi\alpha_{\mathrm{H}}c^2}{m_{\mathrm{ISCO}}^2}.   
\end{equation}
As the gravitational wave power is nearly Planckian whereas the graviton power is ``cut" at a mass larger than the Planck mass, the inequality $P_{\mathrm{GW}}^{\mathrm{ISCO}}>P_{\mathrm{gr}}^{\mathrm{ISCO}}$ always holds. It approaches equality only if the mass is small at the ISCO. 

If $R_0>R_1$, the final power in gravitational waves vanishes while the final power in gravitons diverges (or becomes Planckian if one introduces a regulator in the final stages of the evaporation process). It is then ensured that $P_{\mathrm{gr}}(t_{\mathrm{ev}})\gg P_{\mathrm{GW}}({t_{\mathrm{ev}}})$. 

This completely determines the powers at the end of the process.\\

On the other hand, the initial power radiated as gravitational waves and as gravitons are given by
\begin{equation}
    P_{\mathrm{GW}}^0 \equiv P_{0} = \frac{64}{5}\frac{G^4m_{0}^5}{c^5R_{0}^5}
\qq{and}
    P_{\mathrm{gr}}^0 =\frac{\xi\alpha_{\mathrm{H}}c^2}{m_{0}^2}.
\end{equation}
This defines a critical radius
\begin{equation}
    R_{\mathrm{G}} \equiv \left(\frac{64}{5}\right)^{1/5}\left(\frac{G^4}{\xi\alpha_{\mathrm{H}}c^7}\right)^{1/5}m_{0}^{7/5},
    \label{eq:R_crit_P}
\end{equation}
such that if $R_0<R_{\mathrm{G}}$, then $P_{\mathrm{GW}}^0>P_{\mathrm{gr}}^0$ while if $R_0>R_{\mathrm{G}}$, then $P_{\mathrm{GW}}^0<P_{\mathrm{gr}}^0$. This can be numerically expressed as 
\begin{equation}
    R_{\mathrm{G}} = \SI{8.4e-4}{Mpc} \left(\frac{m_{0}}{M_{\odot}}\right)^{7/5} = \SI{6.2e-7}{m} \left(\frac{m_{0}}{M_{*}}\right)^{3/2}.
\end{equation}

At this stage, the initial and final conditions are known. This, however, immediately raises a question: how does $R_{\mathrm{G}}$ compare with $R_1$? Interestingly, they do not coincide in general, which means that the initial domination of powers does not determine the final one -- although they are obviously {\it not} uncorrelated. As expected, the mass dependence in both the formulae of $R_{\mathrm{G}}$ and $R_{1}$ exhibit very close but  not exactly identical exponents: the scaling appears with a power 1.5 for $R_1$, and with power $1.4$ for $R_{\mathrm{G}}$. This is not so surprising but not {\it a priori} obvious. The numerical evaluation shows that both radii coincide for $M\sim 1000$ $M_{\mathrm{Pl}}$ where $M_{\mathrm{Pl}}$ is the Planck mass. The case $R_{\mathrm{G}} > R_1$ is therefore mostly formal and corresponds to a physical situation which, although possible, remains unlikely for phenomenological purposes.\\

To fully understand the situation, it is interesting to introduce the following mass:
\begin{equation}
   m_{1} \equiv \left(\frac{45}{256}\right)^{5/2}\left(\frac{64}{5}\right)^{2}\left(\frac{G\alpha_{\mathrm{H}}}{\xi^{4}c^{3}}\right)^{1/2}.
\label{eq:m1}
\end{equation}
 It is such that  $m_0 > m_1 \Longleftrightarrow R_{\mathrm{G}} < R_1$. This allows a refinement of the previous taxonomy.

\begin{itemize}
    \item If $m_0>m_1$:
    \begin{itemize}
        \item if $R_0<R_{\mathrm{G}}$, gravitational wave power dominates at the initial and final stages;
        \item if $R_{\mathrm{G}}<R_0<R_1$, graviton power initially dominates but gravitational wave power finally dominates;
        \item if $R_0>R_1$ graviton power dominates at the initial and final stages.
    \end{itemize}
    \item If $m_0<m_1$:
    \begin{itemize}
        \item if $R_0<R_1$, gravitational wave power dominates at the initial and final stages;
        \item if $R_1<R_0<R_{\mathrm{G}}$, gravitational wave power initially dominates but graviton power finally dominates;
        \item if $R_0>R_{\mathrm{G}}$, graviton power dominates at the initial and final stages.
    \end{itemize}
\end{itemize}

\begin{figure}[t!]
    \centering
    \includegraphics[width=0.45\textwidth]{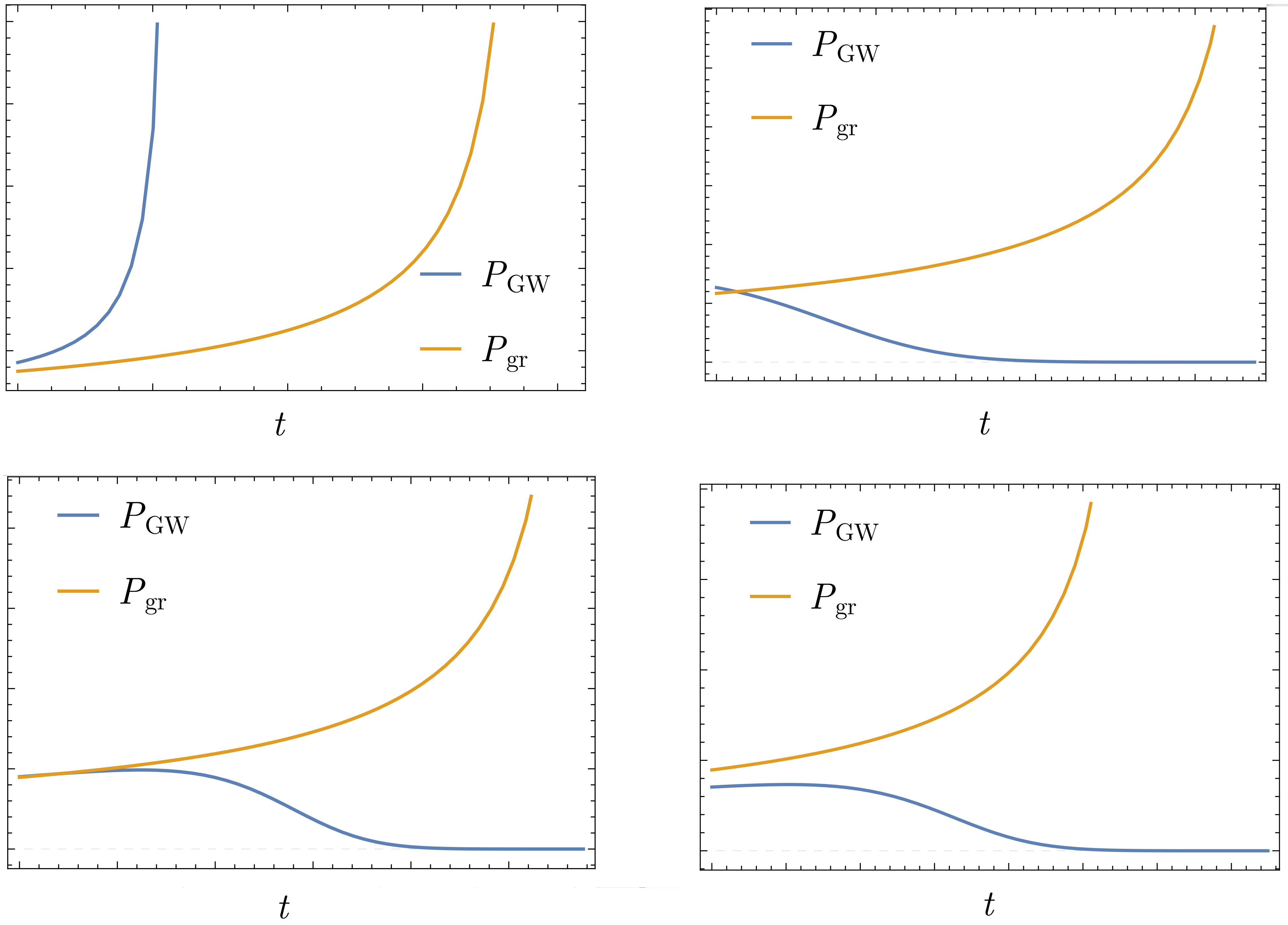}
    \caption{Some sketches of the typical time evolution (from $0$ to $t_{\mathrm{ev}}$) of the gravitational wave radiated power $P_{\mathrm{GW}}$ and its graviton counterpart $P_{\mathrm{gr}}$, showing the diversity of situations and illustrating the taxonomy presented in the text. The top left plot depicts an outspiralling system, with  $P_{\mathrm{GW}}$ dominating at the initial and final stages. The top right and bottom left plots depict situations where $P_{\mathrm{GW}}$ and $P_{\mathrm{gr}}$ intersect, respectively in the outspiralling regime and in the non-monotonic regime. Eventually, the bottom right plot represents an outspiralling system for which $P_{\mathrm{gr}}$ dominates at all times.}
    \label{fig:Typical_Power}
\end{figure}

\begin{figure}
    \centering
    \includegraphics[width=0.45\textwidth]{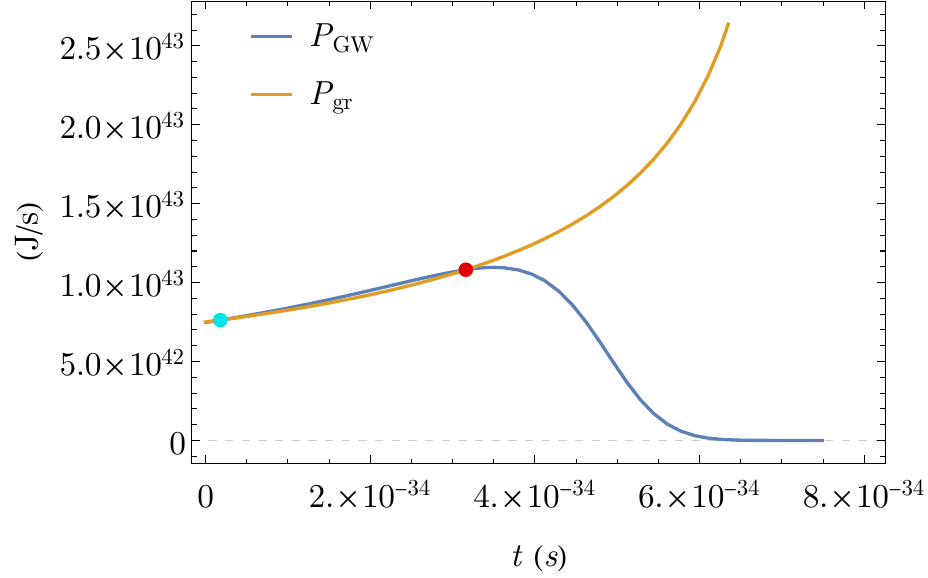}
    \caption{Atypical situation where, initially, $P_{\mathrm{gr}} > P_{\mathrm{GW}}$ and the powers exhibit two points of intersection. This only occurs in the case where the initial conditions are extremely close to the critical values}
    \label{fig:Two_Intersections}
\end{figure}

Figure \ref{fig:Typical_Power} illustrates some typical behaviors. Although it is not of great phenomenological relevance, a subtlety should be underlined at this stage: one might assume that if either gravitational wave or graviton power dominates both at the initial and final stages, it should always dominate. This is however not the case. Figure \ref{fig:Two_Intersections} exhibits a double crossing situation where graviton power dominates at the beginning and at the end but {\it not} during the entire dynamics. In practice, this however happens only for a tiny portion of the parameter space, when the parameters are fine-tuned extremely close to the bounds.

As a conclusion, except for exceptional cases, the preceding list captures the diversity of behaviors for both powers as a function of the initial conditions. Obviously, the answer to the competition between gravitons and gravitational waves as the main source of space-time vibration did not have a simple intuitive answer and required a somehow extensive investigation.

\subsection{Radiated energy}

\subsubsection{Comparison between gravitational waves and gravitons}

Beyond the comparison of instantaneous powers, it makes sense to evaluate the full amount of energy radiated as gravitons and as gravitational waves during the entire process from the remote past to the ISCO.\\

\begin{figure}[htb]
    \centering
    \includegraphics[width=0.49\textwidth]{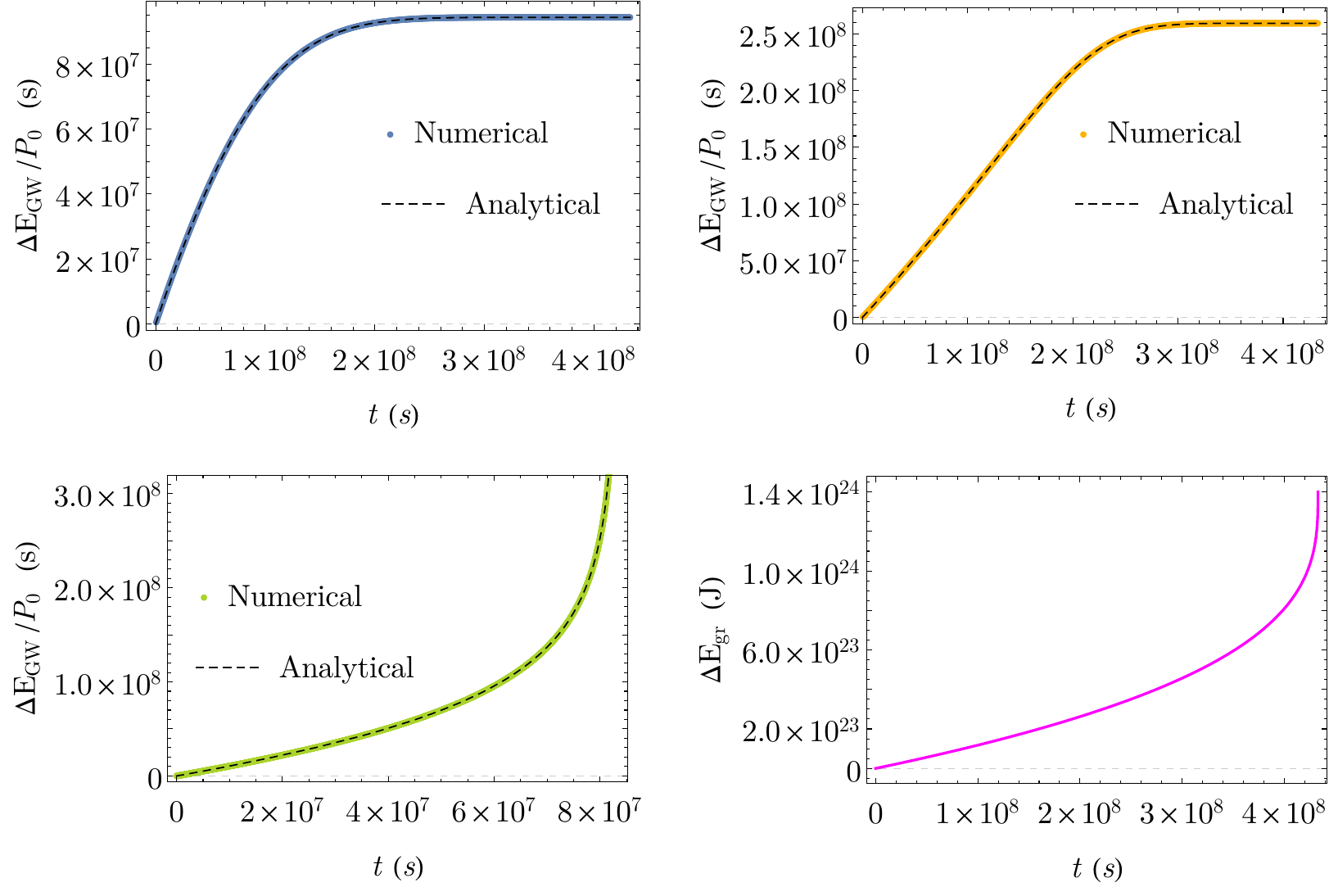}
    \caption{On the top and bottom left plots, the temporal evolution of the total energy radiated as gravitational waves (divided by the initial value of the power $P_{0}$) is depicted in each of the three regimes: blue for outspiralling, green for inspiralling, and orange for the intermediate case. These curves correspond to the results from the numerical integration of Eq. \eqref{eq:Energy_gw} while the black dashed ones correspond to analytical formulas given by Eq. \eqref{eq:E_gw_analytical_outspiral} and Eq. \eqref{eq:E_gw_analytical_inspiral}. The bottom right plot displays the radiated energy as gravitons given in Eq. \eqref{eq:E_gr}.}
    \label{fig:Radiated_Energy_GW}
\end{figure}

\begin{figure}
    \centering
    \includegraphics[width=0.45\textwidth]{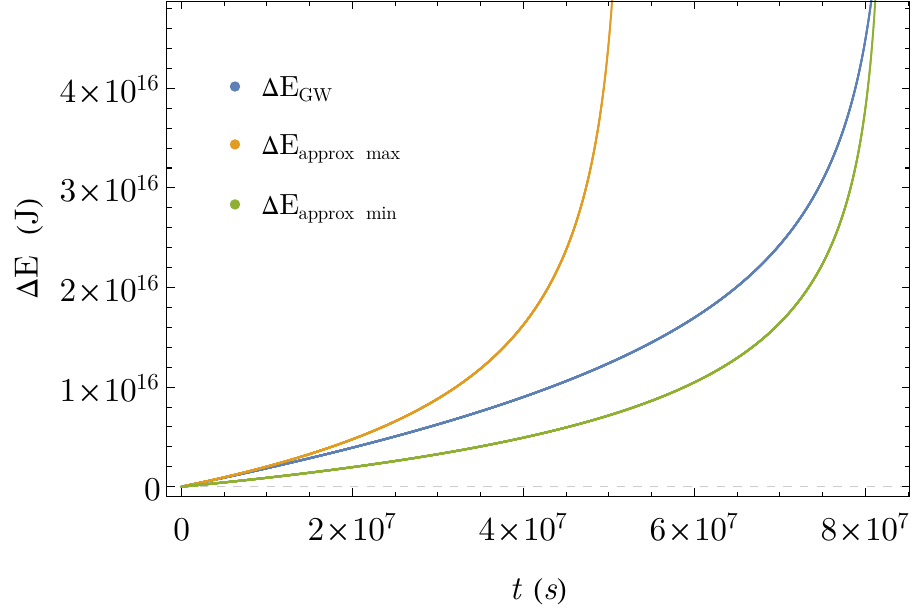}
    \caption{Evolution of the total energy radiated as gravitational waves for the actual system with a varying mass (in blue) together with its approximation by either a constant mass equal to the initial mass (denoted by $\Delta E_{\mathrm{approx max}}$ and represented by the orange curve) or by a constant mass equal to the final one (denoted by $\Delta E_{\mathrm{approx min}}$ and represented by the green curve).}
    \label{fig:energy_approx}
\end{figure}

The energy emitted as gravitons is simply given by $\Delta E_{\mathrm{gr}}=\xi\Delta m c^2$ which leads to

\begin{equation}
   \Delta E_{\mathrm{gr}}=\xi m_0 c^2 \left( 1-\left(1-\frac{t}{t_{\mathrm{ev}}}\right)^{\frac{1}{3}} \right).
\label{eq:E_gr}
\end{equation}
The typical behavior is shown in the bottom right plot of Fig. \ref{fig:Radiated_Energy_GW}. The energy emitted as gravitational waves is obviously less simple to evaluate as soon as the mass varies. Once again, for $t$ in the interval $[t_0, t_{\mathrm{ISCO}}]$ we define
\begin{equation}
    \Delta E_{\mathrm{GW}} = \int_{t_0}^{t} P_{\mathrm{GW}}(t') \dd{t'},
    \label{eq:Energy_gw}
\end{equation}
with $P_{\mathrm{GW}}$ given by Eq. \eqref{eq:Power_gw}. 
The three regimes associated with the behavior of $P_{\mathrm{GW}}$ are present for $\Delta E_{\mathrm{GW}}$ in Fig. \ref{fig:Radiated_Energy_GW}. When the system outspirals, the total energy rapidly reaches a nearly constant value as the radiated power become negligible (both the increase of the orbital separation and the mass loss tend to decrease $P_{\mathrm{GW}}$). When the system inspirals, one recovers the usual trend but with slight differences. Figure \ref{fig:energy_approx} shows the evolution of the total energy radiated as gravitational waves for the actual system with a varying mass together with its approximation by either a constant mass equal to the initial mass or by a constant mass equal to the final one. As expected, the actual behavior lies in between. \\

This raises an interesting comment. We have seen previously that the gravitational wave power necessarily increases when the system inspirals (except, in some very specific situations, for a brief amount of time). This was not obvious as the mass loss competes with the decrease of the orbital separation. We now ask a different question: what about the total energy emitted as gravitational waves? Let us assume that the mass of the BHs varies from $m_0$ initially to $m_{\mathrm{f}}$ at the ISCO. Is $\Delta E_{\mathrm{GW}}$ better approximated by a constant mass equal to $m_0$ of to $m_{\mathrm{f}}$? Otherwise stated: which part of the dynamics does release more energy in the surrounding space? Even if the final power dominates it could be that the time integration favors the initial stages where the system spends more time (in addition of being more massive). We do not address this question frontally and the main conclusion will be given at the end of the point 3 of this subsection. \\

To conclude the present subsection, let us make a brief digression that might help the reader wishing to use explicit formulas. It is indeed possible to compute analytically the integral given by Eq. \eqref{eq:Energy_gw} once the expression of the radiated power $P_{\mathrm{GW}}$ given by Eq. \eqref{eq:Power_gw} is plugged into it. To this purpose, we recognize incomplete Beta functions
\begin{equation}
    B(z ; a, b) \equiv \int_{0}^{z} u^{a-1} (1 - u)^{b-1} \dd{u}
\label{eq:beta_func}
\end{equation}
by performing simple changes of variables. For this to work, one quickly finds out that certain mathematical manipulations are allowed if and only if one specifies in which regime -- whether outspiralling or inspiralling -- we are, which provides a constraint on the sign of the term $1 - \frac{t_{\mathrm{ev}}}{6t_{\mathrm{cc}}}$. Consequently, we expect two different expressions for the integrated power depending on the regime under consideration, which reflects the fact that $\Delta E_{\mathrm{GW}}$ inherits from the behaviour of $P_{\mathrm{GW}}$. Furthermore, for most purposes, it is generally easier to manipulate hypergeometric functions rather than incomplete Beta functions. The link between them, more specifically between $B(z;a,b)$ and the Gaussian (or ordinary) hypergeometric function ${}_{2}F_{1}(a,b,c;z)$ is ensured by the following relation \cite{gradshteyn2007}
\begin{equation}
     B(z;a,b) = \frac{z^{a}}{a} {}_{2}F_{1}(a,1-b, 1+a;z).
\label{eq:beta_and_hypergeo}
\end{equation}

Therefore, in the \textit{outspiralling} case, the change of variable $u = \left(1 - t'/t_{\mathrm{ev}}\right)^{6}$ leads to 
\begin{equation}
\begin{split}
    &\Delta E_{\mathrm{GW}} = \frac{18}{123} \frac{P_{0} t_{\mathrm{cc}}}{\left(1 - \frac{t_{\mathrm{ev}}}{6 t_{\mathrm{cc}}}\right)^{\frac{1}{4}}} \Bigg[ {}_{2}F_{1}\left(\frac{23}{18}, \frac{5}{4}, \frac{41}{18} ; \frac{t_{\mathrm{ev}}}{t_{\mathrm{ev}} - 6 t_{\mathrm{cc}}}\right) \\ &- \left(1- \frac{t}{t_{\mathrm{ev}}}\right)^{\frac{23}{3}} {}_{2}F_{1}\left(\frac{23}{18}, \frac{5}{4}; \frac{41}{18} ; \frac{t_{\mathrm{ev}}}{t_{\mathrm{ev}} - 6 t_{\mathrm{cc}}} \left(1- \frac{t}{t_{\mathrm{ev}}}\right)^{6}\right) \Bigg]
\label{eq:E_gw_analytical_outspiral}
\end{split}
\end{equation}
whereas in the \textit{inspiralling} case, the change of variable $u = \frac{t_{\mathrm{ev}}}{6t_{\mathrm{cc}}} \left[ \left(1- \frac{t'}{t_{\mathrm{ev}}}\right)^{6} - 1\right] +1$ provides the following analytical result:
\begin{widetext}
\begin{equation}
\begin{split}
    \Delta E_{\mathrm{GW}} = &4P_{0} t_{\mathrm{cc}}\left(1 - \frac{6 t_{\mathrm{cc}}}{t_{\mathrm{ev}}}\right)^{5/18} \Bigg[ - \left(1- \frac{t}{t_{\mathrm{ev}}}\right)^{23/3} {}_{2}F_{1}\left(-\frac{1}{4}, -\frac{5}{18}, \frac{3}{4} ; \frac{1}{1-\frac{t_{\mathrm{ev}}}{6 t_{\mathrm{cc}}}} \right) \\ &\left( \frac{t_{\mathrm{ev}}}{6 t_{\mathrm{cc}}}\left[\left(1 - \frac{t}{t_{\mathrm{ev}}}\right)^{6}-1 \right] +1 \right)^{-1/4} {}_{2}F_{1}\left(-\frac{1}{4}, -\frac{5}{18}, \frac{3}{4} ; \frac{1}{\frac{6 t_{\mathrm{cc}}}{t_{\mathrm{ev}}} - 1} \left[\left(1 - \frac{t}{t_{\mathrm{ev}}}\right)^{6}-1 \right] + \frac{1}{1-\frac{t_{\mathrm{ev}}}{6 t_{\mathrm{cc}}}}\right) \Bigg].
\label{eq:E_gw_analytical_inspiral}
\end{split}
\end{equation}
\end{widetext}
On the top plots and the bottom left plot of Fig. \ref{fig:Radiated_Energy_GW}, it is shown that these analytical formulas fit perfectly the curves resulting from the numerical integration of Eq. \eqref{eq:Energy_gw}. Using the main properties of hypergeometric functions, one can recover the main characteristic behaviors exhibited in Fig. \ref{fig:Radiated_Energy_GW}, for instance the convergence of $ \Delta E_{\mathrm{GW}}$ to a finite value for $t \to t_{\mathrm{ev}}$ in the outspiralling case using that $\lim_{z \to 0} {}_{2}F_{1}(a,b,c;z) = 1$.

\subsubsection{Fully time-integrated process}

In the previous analysis, we focused on a process assumed to take place between an arbitrary initial time and a final moment defined as the merging (or ISCO) time. This led to an impressive diversity of conclusions. It is however also possible to consider the fully time-integrated process without interrupting the reasoning at merging. This leads to a (maybe surprising) generic result: the total energy released as spacetime vibrations by gravitational ways is always of the same order (or smaller) than the energy released as gravitons.\\

The total energy radiated as gravitational waves between a remote initial time and the ISCO is, if the mass does not vary substantially,

\begin{equation}
    \Delta E_{\mathrm{GW}}=\frac{1}{24}m_0c^2.
    \label{eq:delta1}
\end{equation}

On the other hand, the total energy emitted as gravitons over an unlimited amount of time is trivially

\begin{equation}
    \Delta E_{\mathrm{gr}}^{\mathrm{full}}=2\xi m_0c^2,
    \label{eq:deltagrav}
\end{equation}
where the superscript ``full" means here that we also consider the evaporation even after merging. If the black hole is hot enough to emit all standard model degrees of freedom, $\xi=1/64$. This elegantly leads to

\begin{equation}
    \Delta E_{\mathrm{GW}}\sim \Delta E_{\mathrm{gr}}^{\mathrm{full}} ,
    \label{eq:deltagrav}
\end{equation}
meaning that as soon as the dynamics of the system is such that it inspirals  -- that is as soon as the evaporation is not too fast at the initial stage -- the gravitational wave energy and the graviton energy are of the same order. If the system outspirals, gravitons obviously dominate strongly.\\

Let us now question the validity of this result by considering several subtleties. 

First, the energy emitted during the merging and ringdown phase can, obviously, not be captured by Eq. \eqref{eq:delta1} which is simply obtained by evaluating the Newtonian orbital energy at the ISCO. The measurements \cite{LIGOScientific:2016aoc} however suggest that the ``missing mass" between the initial black holes and the final one is of the same order as $\Delta E_{\mathrm{GW}}$. As a crude approximation, one therefore gets

\begin{equation}
    \Delta E^{\mathrm{full}}_{\mathrm{GW}}\sim \frac{1}{12}m_0c^2,
    \label{eq:delta2}
\end{equation}
where the superscript ``full" means that the gravitational waves emitted during the non-perturbative phase are also accounted for. 
This clearly increases the amount of GW energy but does not change the orders of magnitude.

Second, one could instead consider very large black holes with a temperature (during most of the process) smaller than the mass of all massive particles. In that case, $\xi=1/2$. This increases the number of gravitons but does not radically alter the conclusion. Anyway, the energy released as gravitons is of the same order as the energy released as gravitational waves and one has:

\begin{equation}
    \Delta E_{\mathrm{GW}}^{\mathrm{full}} \lesssim \Delta E_{\mathrm{gr}}^{\mathrm{full}}.
    \label{eq:deltagrav}
\end{equation}

Third, one can question the fact that the gravitational wave energy is here approximated assuming a constant mass. As we shall see later, there indeed exist very special cases where the system inspirals with a strongly varying mass. Then, gravitational waves are suppressed and $\Delta E_{\mathrm{GW}}<\Delta E_{\mathrm{gr}}$. This however represents a tiny part of the parameter space requiring strong fine-tuning. For the very vast majority of initial conditions, $\Delta E_{\mathrm{GW}}\sim \Delta E_{\mathrm{gr}}$.

Although mathematically very simple, the fact that both energies are comparable was not {\it a priori} so obvious. A binary system of black holes basically ``excites" spacetime as much through gravitational waves than through gravitons.\\

The only important case where a strong hierarchy appears is when the mass is small enough (or the orbital separation is large enough). If it is such that the system outspirals, the amount of gravitational waves becomes obviously negligible. The very specific situation where the system still merges but with a mass at the ISCO which is much smaller than the initial mass will be analytically studied in the next section.\\

In conclusion, the energy released by gravitational waves is always smaller than the one emitted as gravitons. The bound is saturated for an inspiralling system with slowly varying mass and the hierarchy becomes large if the mass loss becomes sizable during the process. It should be noticed that the amount of emitted gravitons is always (nearly) the same, it is only the power radiated as gravitational waves which highly depends upon the orbit.

\subsubsection{Extreme case}

The  analytical solutions are textbook when the mass is nearly constant. There is however another case, on the extreme other side, for which a simple analytical solution can be obtained. It can indeed be shown \cite{Blachier:2023ygh} that a system whose full evaporation time is exactly equal to the merging time does actually exist (which is not obvious as it might have been guessed to outspiral) and is easy to solve. In this situation,  the power radiated by the emission of gravitational waves is given by
\begin{equation}
    P_{\mathrm{GW}} = P_0\left(1- \frac{t}{t_{\mathrm{ev}}}\right)^{-\frac{5}{6}}.
\end{equation}
Using Eq. \eqref{eq:Energy_gw}, a simple analytic expression for the energy dissipated can be found:
\begin{widetext}
\begin{equation}
    \Delta E_{\mathrm{GW}} =  \left(\frac{128}{5}\right)\left(\frac{45}{256}\right)^{\frac{5}{4}}c^{\frac{5}{4}}\alpha_{\mathrm{H}}^{\frac{1}{4}}G^{\frac{1}{4}}m_0^{\frac{1}{2}}\left(1-\left[1-\frac{t}{t_{\mathrm{ev}}}\right]^{\frac{1}{6}}\right).
\end{equation}
\end{widetext}
This allows to evaluate the ratio of emitted energies at the ISCO:
\begin{equation}
    \frac{\Delta  E_{\mathrm{GW}}}{\Delta E_{\mathrm{gr}}}(t_{\mathrm{ISCO}}) = \frac{\left(\frac{128}{5}\right)\left(\frac{45}{256}\right)^{\frac{5}{4}}\alpha_{\mathrm{H}}^{\frac{1}{4}}G^{\frac{1}{4}}}{\xi m_0^{\frac{1}{2}}c^{\frac{3}{4}} + 12\left(\frac{45}{256}\right)^{\frac{1}{4}}G^{\frac{1}{4}}\xi\alpha_{\mathrm{H}}^{\frac{1}{4}}}.
    \label{deltaEextreme}
\end{equation}

Interestingly, this shows that as soon as the initial mass is not negligible -- if it was the case, it would anyway mean that the process would be nearly instantaneous -- the gravitational wave energy is highly suppressed when compared to the one associated with gravitons. Although quite expected on the one hand, this is, on the other hand, a non-trivial result as the gravitational wave power emitted at the ISCO is still Planckian. 

In the extreme case considered here, the BHs have nearly fully evaporated when they reach the ISCO. This implies that $\Delta E_{\mathrm{gr}}\approx\xi m_0c^2$. In more details, Eq. (\ref{deltaEextreme}) therefore shows that

\begin{equation}
    \Delta  E_{\mathrm{GW}}\propto \sqrt{m_0c^2}.
\end{equation}
This answers the question that we previously asked. When the mass loss is substantial during the merging process, the total amount of energy emitted as gravitational waves is neither approximated by assuming a constant mass equal to the initial one nor a constant mass equal to the final one.  

\section{Space expansion}

We now focus on a second situation where a competition takes place between the dynamics driven by the classical BH behavior and the evolution due to Hawking evaporation. Let us consider a gas of BHs in the early universe. Those BHs evaporate and produce radiation. On the one hand, the expansion of space tends to favor matter -- that is BHs -- over radiation as the energy density scales as $a^{-3}$ for the former and as $a^{-4}$ for the latter, $a$ being the scale factor. On the other hand, the evaporation process tends to favor radiation over matter as it converts the BH mass into relativistic quanta. It is not {\it a priori} obvious to determine the evolution of the system. Is the energy density of the universe dominated by matter or radiation? If a transition takes place, when does this happen and why?

In most previous studies, authors  focused mostly on the ratio between the BH energy density and the total energy density (thus comprising both matter and radiation) \cite{Cheek:2022mmy, Dienes:2022zgd}, which somehow blurs the competitive effects between radiation due to the evaporation and the contribution of BHs themselves. The first articles to address the question analytically were \cite{Barrow:1991dn, Barrow:1992hq}. In \cite{Barrow:1991dn}, the focus was on a power law mass spectrum for BHs while we will consider here  the monochromatic case, as it makes the situation much clearer. We provide analytical solutions, completing and refining the works of \cite{Gutierrez:2017ibk, Barrow:1992hq} as well as correcting certain mistakes in \cite{Barrow:1992hq}.

\subsection{Numerical results}

The gas of BHs is modeled as a dust-like perfect fluid ({\it i.e.} $p_{\mathrm{m}} = 0$) whose constituents are Schwarzschild black holes, with an energy density $\rho_{\mathrm{m}}$. If we denote by $f_{\mathrm{PBH}}(m, t)$ the distribution of BHs of mass $m$ at time $t$ -- the so-called \textit{mass spectrum} -- the BH energy density is given by
\begin{equation}
    \rho_{\mathrm{m}}(t) = \int_{0}^{\infty} f_{\mathrm{PBH}}(m, t) m \dd{m},
\label{eq:rhoPBH_def}
\end{equation}

\begin{figure}
    \centering
    \includegraphics[width = 0.45 \textwidth]{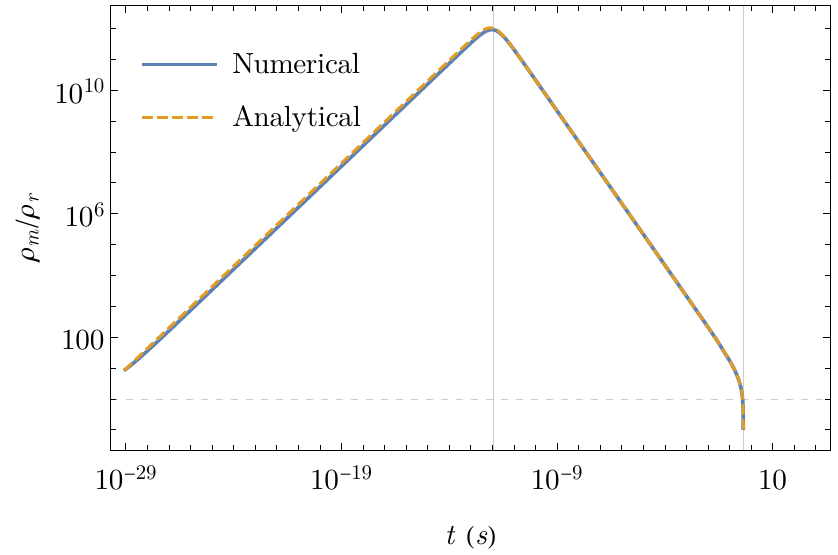}	
    \caption{Ratio of matter over radiation energy densities $y=\rho_{\mathrm{m}}/\rho_{\mathrm{r}}$, as a function of time, from an initial formation time $t_{0}$ to the time of evaporation $t_{\mathrm{ev}}$. The blue curve represents the results from numerical integration while the orange dotted one is our analytical solution. The dashed horizontal line corresponds to $y=1$ while the two vertical lines respectively correspond to $t= t_{\mathrm{inv}}$ and $t = t_{\mathrm{ev}}$.}
    \label{fig:rhoPBH/rhoR}
\end{figure}

The energy density for radiation is denoted by $\rho_{\mathrm{r}}$. The initial total density is $\rho_{0}$ and the parameter $\beta$ fixes the initial ratio between matter and radiation, namely $\rho_{\mathrm{m}}(t_{0}) = \beta \rho_{0}$ and $\rho_{\mathrm{r}}(t_{0}) = (1-\beta) \rho_{0}$. For an evaporating distribution of BHs, the continuity equations are modified by an interaction term (see for instance \cite{Gutierrez:2017ibk}), leading to the so-called Friedmann-Boltzmann equations. In the case of a purely monochromatic spectrum, we thus have the following set of coupled ordinary differential equations for the energy density of BHs (thus matter) and radiation $\rho_{\mathrm{r}}$:
\begin{equation}
    \dot{\rho}_{\mathrm{m}} + 3H \rho_{\mathrm{m}} = \rho_{\mathrm{m}} \frac{\dot{m}}{m} \qq{and} \dot{\rho}_{\mathrm{r}} + 4H \rho_{\mathrm{r}} =  -\rho_{\mathrm{m}} \frac{\dot{m}}{m},
\label{eq:ODE}
\end{equation}
supplemented by the Friedmann equation for two components universe:
\begin{equation}
    H^2 = \frac{8\pi G}{3c^{2}} \left(\rho_{\mathrm{m}} + \rho_{\mathrm{r}}\right).
\end{equation}
This set, knowing the evolution law for $m$, as given by Eq. \eqref{eq:hawking_mass}, can easily be numerically integrated. The result is displayed by the blue curve in Fig. \ref{fig:rhoPBH/rhoR} for $\beta = 0.9$. The analytical solution discussed in the following subsections is represented in dashed orange. Several interesting features should be noticed. \\

First, the ratio $\rho_{\mathrm{m}}/\rho_{\mathrm{r}}$ is initially increasing. This means that, at the beginning of the process, the differential dilution (favouring matter over radiation) is more efficient  than Hawking evaporation. Although BHs lose mass, their contribution to the energy budget of the Universe increases with time. 

Second, there exists an inflection point -- corresponding to the first vertical line and defining the time $t_{\mathrm{inv}}$ -- where the derivative of $\rho_{\mathrm{m}}/\rho_{\mathrm{r}}$ vanishes. It means that at some point, the evaporation becomes efficient enough to overcome the dilution. Very interestingly, this does {\it not} happen when the Hawking evaporation becomes deeply explosive, that is when the mass plunges to zero. This can be understood by looking carefully at the mass evolution given by Eq. \eqref{eq:hawking_mass}. It is correct to think that this behavior, driven by a positive feedback and a negative heat capacity, is somehow catastrophic and leads to a singularity (in the sense that $\dd{m}/\dd{t}$ diverges) at $t=t_{\mathrm{ev}}$. However, the actual evaporation time differs only by a factor of $1/3$ from the one naively extrapolated from the initial evaporation. This basically means that contrary to what is usually assumed, the evaporation is {\it not} that slow at the initial time (when compared to the actual evaporation time).

Third, although both the previously mentioned regimes are obviously described by power laws, the absolute value of the slope of the second one is larger. When the evaporation process begins to dominate, it does so more efficiently. Let us underline that by evaporation ``domination", we mean here that the derivative of $\rho_{\mathrm{m}}/\rho_{\mathrm{r}}$ is negative. However, during nearly all the evolution (except, of course, if $\beta$ is chosen tiny), matter dominates and the scale factor behaves as $a\propto t^{2/3}$.

Fourth, there exists a final regime -- corresponding to the second vertical line -- where the considered ratio decreases very fast before vanishing. This corresponds to the ``explosion" of the BHs. For most -- but not all -- of the parameter space, this happens when $\rho_{\mathrm{m}}/\rho_{\mathrm{r}}$ becomes close to its initial value.

Fifth, one should also be careful not to be fooled by the logarithmic scale used in Fig. \ref{fig:rhoPBH/rhoR}: in cosmic time, the inflection point happens very soon after the beginning of the process. However, $\rho_m/\rho_r$ has already considerably increased in this small amount of time.

%Two vertical lines have been plotted, as well as one dashed horizontal line corresponding to $\rho_{\mathrm{m}}/\rho_{\mathrm{r}} = 1$. The last vertical line corresponds to $t= t_{\mathrm{ev}}$ and matches the moment where Hawking evaporation is the most efficient, and correspondingly, the moment for which $\rho_{\mathrm{r}} > \rho_{\mathrm{m}}$. The first vertical line is however more surprising.

\subsection{Analytical solutions}

\begin{figure}[h]
    \centering
    \includegraphics[width = 0.43 \textwidth]{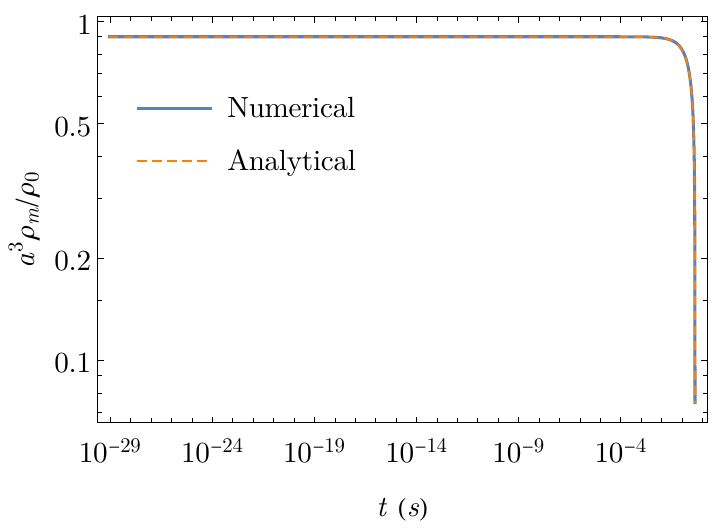}
    \caption{Evolution of the comoving matter energy density (normalized to the reference energy scale $\rho_{0}$) as a function of time from $t_{0}$ to $t_{\mathrm{ev}}$, with the same color codes as in Fig. \ref{fig:rhoPBH/rhoR}.}
    \label{fig:rhoPBH}
\end{figure}

In the case of a purely monochromatic spectrum, $f_{\mathrm{PBH}}$ remains monochromatic at all times so that Eq. \eqref{eq:rhoPBH_def} straightforwardly becomes
\begin{equation}
    \rho_{\mathrm{m}}(t) = \rho_{\mathrm{m}}^{0} \left(\frac{a(t_{0})}{a(t)}\right)^{3} \left(1 - \frac{t}{t_{\mathrm{ev}}}\right)^{\frac{1}{3}},
\label{eq:rhoPBH_analytical}
\end{equation}
with $\rho_{\mathrm{m}}^{0} \equiv \beta \rho_{0} \left(1-\frac{t_{0}}{t_{\mathrm{ev}}}\right)^{-1/3}$. As displayed in Fig. \ref{fig:rhoPBH} and as expected, the analytical solution for $ \rho_{\mathrm{m}}$ matches perfectly the one resulting from the numerical integration. 

\begin{figure}[h]
    \centering
    \includegraphics[width = 0.45 \textwidth]{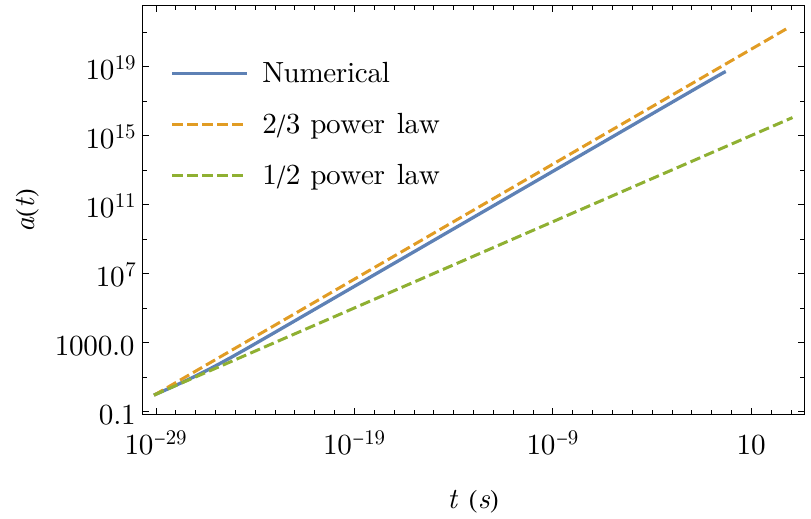}
    \caption{Scale factor $a(t)$ as a function of time, resulting from numerical integration and represented in blue. For comparison, in dashed orange (resp. green), is represented $a(t) \propto t^{2/3}$ (resp. $a(t) \propto t^{1/2}$). Even for $\beta = 0.1$, which is the value chosen for this plot, it can be seen that the slope 2/3 is quickly recovered. As expected, the fit  $a(t) \propto t^{2/3}$ is increasingly better as $\beta$ approaches unity.}
    \label{fig:scale_factor}
\end{figure}

More interestingly, an approaching analytical formula can also be derived for $\rho_{\mathrm{r}}$. As shown in Eq. \eqref{eq:ODE}, it obeys a differential equation of the form $y' + p(x) y = b(x)$ whose solution is given by $y(x) = e^{-P(x)} \int e^{P(s)} b(s) \dd{s} + C e^{-P(x)}$ where $P$ is the primitive of the function $p$ and where $C$ is a constant fixed by specifying initial conditions. In our case, $p(t) = 4 H$ which is straightforwardly integrated into $P(t) = a(t_{0})^{4} / a(t)^{4}$. Furthermore, the right-hand side of the ODE is now completely determined using Eq. \eqref{eq:hawking_mass} and Eq. \eqref{eq:rhoPBH_analytical}. Enforcing that at the initial time $t_{0}$, $\rho_{\mathrm{r}}$ takes the value $(1- \beta) \rho_{0}$, we obtain:
\begin{equation}
\begin{split}
   \rho_{\mathrm{r}}(t) = &\frac{\rho_{\mathrm{m}}^{0}}{3a(t_{0}) t_{\mathrm{ev}}}\left(\frac{a(t_{0})}{a(t)}\right)^{4} \int_{t_{0}}^{t} a(s) \left(1 - \frac{s}{t_{\mathrm{ev}}}\right)^{-\frac{2}{3}} \dd{s} \\ &+ (1- \beta) \rho_{0} \left(\frac{a(t_{0})}{a(t)}\right)^{4}.
\end{split}
\label{eq:interm_rhoR}
\end{equation}
The only difficulty to overcome is to compute the integral involved in the above equation, which necessarily means specifying an expression for the scale factor $a(t)$. At this step, we are coerced into resorting to numerical resolutions of the set of coupled differential equations which show that, as could be expected, $a(t)$ nearly behaves as a power law, but with an exponent that can, in principle, shift between $1/2$ (radiation domination) and $2/3$ (matter domination). Actually, even when radiation dominates initially -- that is to say $\beta$ is close to 0 -- $a(t)$ quickly recovers the matter-dominated trend as it can be seen in Fig. \ref{fig:scale_factor}. As a first approximation, we shall thus consider that $a(t) \approx A t^{2/3}$ with $A$ a constant. This is natural as we have previously shown that $\rho_m/\rho_r$ strongly increases at the beginning of the process and still dominates even when the evaporation begins to become significant. 

\begin{figure}[h]
    \centering
    \includegraphics[width = 0.45 \textwidth]{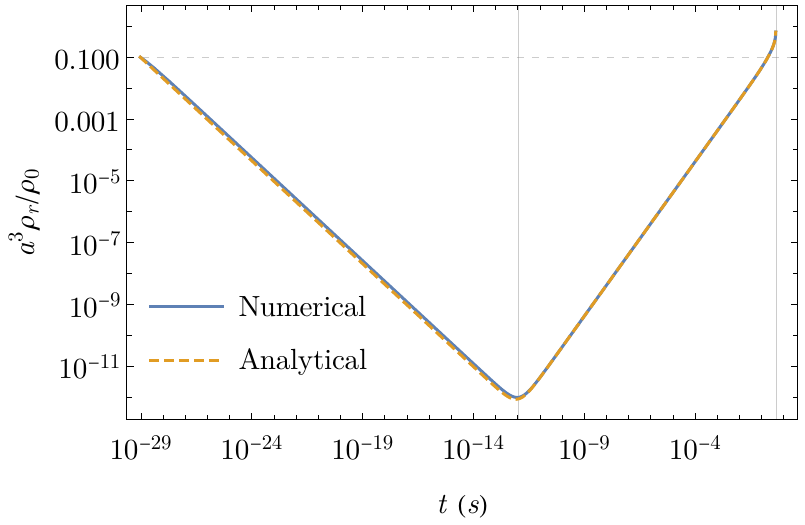}
    \caption{Evolution of the comoving radiation energy density (normalized to the reference energy scale $\rho_{0}$) as a function of time from $t_{0}$ to $t_{\mathrm{ev}}$, with the same color codes as in Fig. \ref{fig:rhoPBH/rhoR}. The horizontal dashed line corresponds to $y = a^{3}(t_{0}) \rho_{\mathrm{r}}(t_{0}) = (1-\beta) \rho_{0}$ while the two vertical lines respectively correspond to $t= t_{\mathrm{inv}}$ and $t = t_{\mathrm{ev}}$.}
    \label{fig:rhoR_analytical}
\end{figure}

In this case, the integral is analytical and can be split into two parts in order to recognize incomplete Beta functions (see Eq. \eqref{eq:beta_func}). Using the fact that they can be expressed in terms of hypergeometric functions (see Eq. \eqref{eq:beta_and_hypergeo}), we ultimately have
\begin{equation}
\begin{split}
    \rho_{\mathrm{r}}(t) = &\frac{A \rho_{\mathrm{m}}^{0}}{5a(t_{0})t_{\mathrm{ev}}}\left(\frac{a(t_{0})}{a(t)}\right)^{4} \Bigg[ t^{5/3}{}_{2}F_{1}\left(\frac{5}{3}, \frac{2}{3}, \frac{8}{3} ; \frac{t}{t_{\mathrm{ev}}} \right) \\ &- t_{0}^{5/3}{}_{2}F_{1}\left(\frac{5}{3}, \frac{2}{3}, \frac{8}{3} ; \frac{t_{0}}{t_{\mathrm{ev}}} \right) \Bigg] + (1-\beta) \rho_{0} \left(\frac{a(t_{0})}{a(t)}\right)^{4}.
\end{split}
\label{eq:rhoR_analytical}
\end{equation}
As it can be seen in Fig. \ref{fig:rhoR_analytical}, the above analytical form matches very well the numerical results. However, one should notice that as $\beta$ gets closer to 0, that is to say as the initial amount of matter becomes negligible, the fit of the first branch of the numerical integration becomes less accurate. This is the direct consequence of having assumed $a(t) \propto t^{2/3}$ during the whole evolution of the system (obviously, when $\beta\sim 0$, the universe is initially radiation-dominated and, at the very beginning of the evolution, $a(t) \propto t^{1/2}$). A possible refinement would consist in finding the value $t_{*}$ for which the power-law exponent for the scale factor shifts from $1/2$ to $2/3$, and splitting the integral in Eq. \eqref{eq:interm_rhoR} in order to plug the correct expressions of $a(t)$ for $t<t_{*}$. However, such manipulations only make the computations heavier,  bringing neither new physics nor substantial improvements in accuracy.\\

Let us analyse Eq. \eqref{eq:rhoR_analytical} in details: the constant term $t_{0}^{5/3}{}_{2}F_{1}\left(\frac{5}{3}, \frac{2}{3}, \frac{8}{3} ; \frac{t_{0}}{t_{\mathrm{ev}}} \right)$ is always negligible when compared to its time-dependent counterpart. This, of course, is not fully correct as the first term in Eq. \eqref{eq:rhoR_analytical} does not vanish initially at $t=t_{0}$. However, when focusing on most of the temporal evolution of $\rho_{\mathrm{r}}$, this approximation is completely legitimate. Furthermore, as the behaviour of the hypergeometric function is such that \footnote{To prove this, one can use the integral representation \begin{equation}
    {}_{2}F_{1}(a,b,c;z) = \frac{\Gamma(c)}{\Gamma(b) \Gamma(c-b)} \int_{0}^{1} \frac{t^{b-1} (1-t)^{c-b-1}}{(1-tz)^{a}} \dd{t}.
\end{equation} Taking the limit $z \to 0$ in the above, and recognizing that the remaining integral is a Beta function which cancels with all of the Gamma functions suffices to establish the result. } $\lim_{z \to 0} {}_{2}F_{1}(a,b,c;z) = 1$, if the system is not close to the typical time of evaporation $t_{\mathrm{ev}}$, it does not contribute. Only for $t \sim t_{\mathrm{ev}}$, does the hypergeometric function play a role, essentially modeling the fact that the BHs have entered the stiff part of the evaporation process, thus leading to a kink in the radiation energy density contribution.

Consequently, for $t<t_{\mathrm{ev}}$, Eq. \eqref{eq:rhoR_analytical} can be simplified to
\begin{equation}
    \rho_{\mathrm{r}}(t) \approx \frac{A \rho_{\mathrm{PBH}}^{0}}{5a(t_{0})t_{\mathrm{ev}}}\left(\frac{a(t_{0})}{a(t)}\right)^{4} t^{\frac{5}{3}} + (1-\beta) \rho_{0} \left(\frac{a(t_{0})}{a(t)}\right)^{4}.
\label{eq:rhoR_analytical_3}
\end{equation}

\begin{figure}[h]
    \centering
    \includegraphics[width = 0.45 \textwidth]{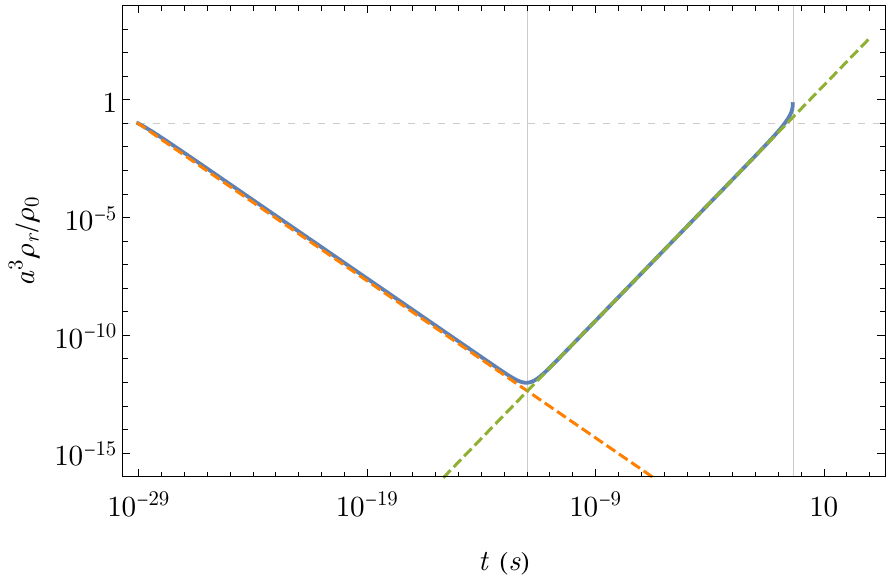}
    \caption{Same plot as in Fig. \ref{fig:rhoR_analytical} but highlighting the different contributing terms of the approximated analytical solution \eqref{eq:rhoR_analytical_3}. The blue curve is the result of the numerical integration. The orange dotted line corresponds to the function $t \mapsto (1-\beta) \left(\frac{a(t_{0})}{a(t)}\right) \sim (1-\beta) t^{-2/3}$ while the green one corresponds to $t \mapsto \frac{A \rho_{\mathrm{PBH}}^{0}}{5a(t)t_{\mathrm{ev}}}t^{5/3} \sim  \frac{A \rho_{\mathrm{PBH}}^{0}}{5t_{\mathrm{ev}}} t$. These two curves exactly intersect at the so-called time of inversion $t_{\mathrm{inv}}$ given by Eq. \eqref{eq:t_inv}.}
    \label{fig:rhoR_simplified}
\end{figure}

As highlighted by Fig. \ref{fig:rhoR_simplified}, the above formula is sufficient to fit the trend of $a^{3}\rho_{\mathrm{r}}$ except for $t$ very close to $t_{\mathrm{ev}}$. It also gives a mathematical explanation of the reason why the curve reaches a minimum \textit{before} reaching the second cutoff associated to the full evaporation as a consequence of the competition between the functions $\frac{A \rho_{\mathrm{PBH}}^{0}}{5a(t_{0})t_{\mathrm{ev}}} t^{5/3}$  and $(1-\beta) \rho_{0}t^{-2/3}$, the second being pure radiation while the first is modeling the smooth evaporation. Both these quantities present monotonic behaviours and the minimum reached by $a^{3}(t) \rho_{\mathrm{r}}(t)$ corresponds to their equality. Equating them straightforwardly gives the so-called time of inversion
\begin{equation}
    t_{\mathrm{inv}} = \left(\frac{1-\beta}{\beta} \frac{5t_{\mathrm{ev}}}{A} \right)^{\frac{3}{5}} \left(1 - \frac{t_{0}}{t_{\mathrm{ev}}} \right)^{\frac{1}{5}}.
\label{eq:t_inv}
\end{equation}
Since $t_{\mathrm{ev}} \propto m_{0}^{3}$, studying $t_{\mathrm{inv}}$ as a function of $t_{\mathrm{ev}}$ shows that, as expected, the smaller the initial mass of black holes, the closer to the initial time of formation $t_{0}$ will $t_{\mathrm{inv}}$ be. In the extreme limit where $\beta = 1$, this competitive effect is not present anymore and the radiation energy density starts by increasing from 0 to a non-vanishing value thanks to the radiation created by the evaporation process.\\

It can also be noticed that for any generic initial condition different from the very specific $\beta=1$ case,  this inversion \textit{always} occurs. This detail was missed in the analysis presented in \cite{Barrow:1992hq} and could be guessed from the plots of \cite{Gutierrez:2017ibk} but was neither explained nor proved. Although it does not affect the big picture of a matter dominated regime followed by a radiation dominated one at the end of the evaporation process, it highlights that even in its early stage, the evaporation process plays a non-negligible role that is not restrained to what occurs near $t_{\mathrm{ev}}$. Furthermore, one could have assumed that for light enough black holes -- that are evaporating fast enough, as $\dd{m}/\dd{t}\propto 1/m^2$ -- radiation immediately dominates. This is not the case. There is no critical initial mass below which this happens. 

Let us finally analyze the end behaviour of $\rho_{\mathrm{r}}$. As expected, it can be seen that it converges to a finite value at $t = t_{\mathrm{ev}}$. Our analytical solutions also exhibit these behaviours, through the fact that $_{2}F_{1}(\frac{5}{3},\frac{2}{3},\frac{8}{3},1) = \Gamma(\frac{1}{3})\Gamma(\frac{8}{3})$. From here, one can plot $\rho_{\mathrm{r}}$ for a range of initial conditions and see that there are some instances, depending upon the initial value of $\beta$, for which the final value of $\rho_{\mathrm{r}}$ at $t_{\mathrm{ev}}$ will exceed the initial value of $\rho_{\mathrm{r}}$, and others such that $\rho_{\mathrm{r}}$ converges to a value that is less than that of the initial conditions. The critical value $\beta_{\mathrm{c}}$ of $\beta$ for which it occurs, is obtained by solving the equation $a^{3}(t_{0}) \rho_{\mathrm{r}}(t_{0}) = a^{3}(t_{\mathrm{ev}}) \rho_{\mathrm{r}}(t_{\mathrm{ev}})$, using Eq. \eqref{fig:rhoR_analytical} and the convergence of the hypergeometric function. It can be well approximated by
\begin{equation}
     \beta_{\mathrm{c}} \approx \frac{1}{\frac{1}{5} \Gamma(\frac{1}{3})\Gamma(\frac{8}{3}) + 1}.
\end{equation}

\subsection{Relics}

In this work, we have resisted the temptation to bring any exotic physics into the game and we stood with purely standard processes. Let us make here an exception. Based on several distinct arguments (see, {\it e.g.}, references in \cite{Barrau:2022eag}), it is possible -- if not probable -- that stable relics are formed at the end of the evaporation process, with masses close to the Planck mass $m_{\mathrm{rel}}\sim m_{\mathrm{Pl}}$. This makes the evolution even more interesting. 

Starting with BHs with masses larger than the Planck mass, the dynamics begin, as usual, by an increase of the matter contribution over the radiation one (in a sense, space expansion overcomes the evaporation). At some point ($t=t_{\mathrm{inv}}$), the trend reverses and the relative contribution of matter begins to decrease (evaporation overcomes space expansion) -- although it remains dominant for a while. At $t\approx t_{\mathrm{ev}}$, the evolution becomes dramatic and $\rho_{\mathrm{m}}/\rho_{\mathrm{r}}$ plunges. If it was not already the case, radiation soon dominates. However, in this case, the matter density does not reach exactly zero as the evaporation stops at $m=m_{\mathrm{rel}}$. Due to the differential dilution, matter (that is relics) contribution slowly begins to increase again. At some point, it eventually drives the cosmological dynamics. Once relics dominate, they obviously keep dominating.

\begin{figure}
    \centering
    \includegraphics[width = 0.45 \textwidth]{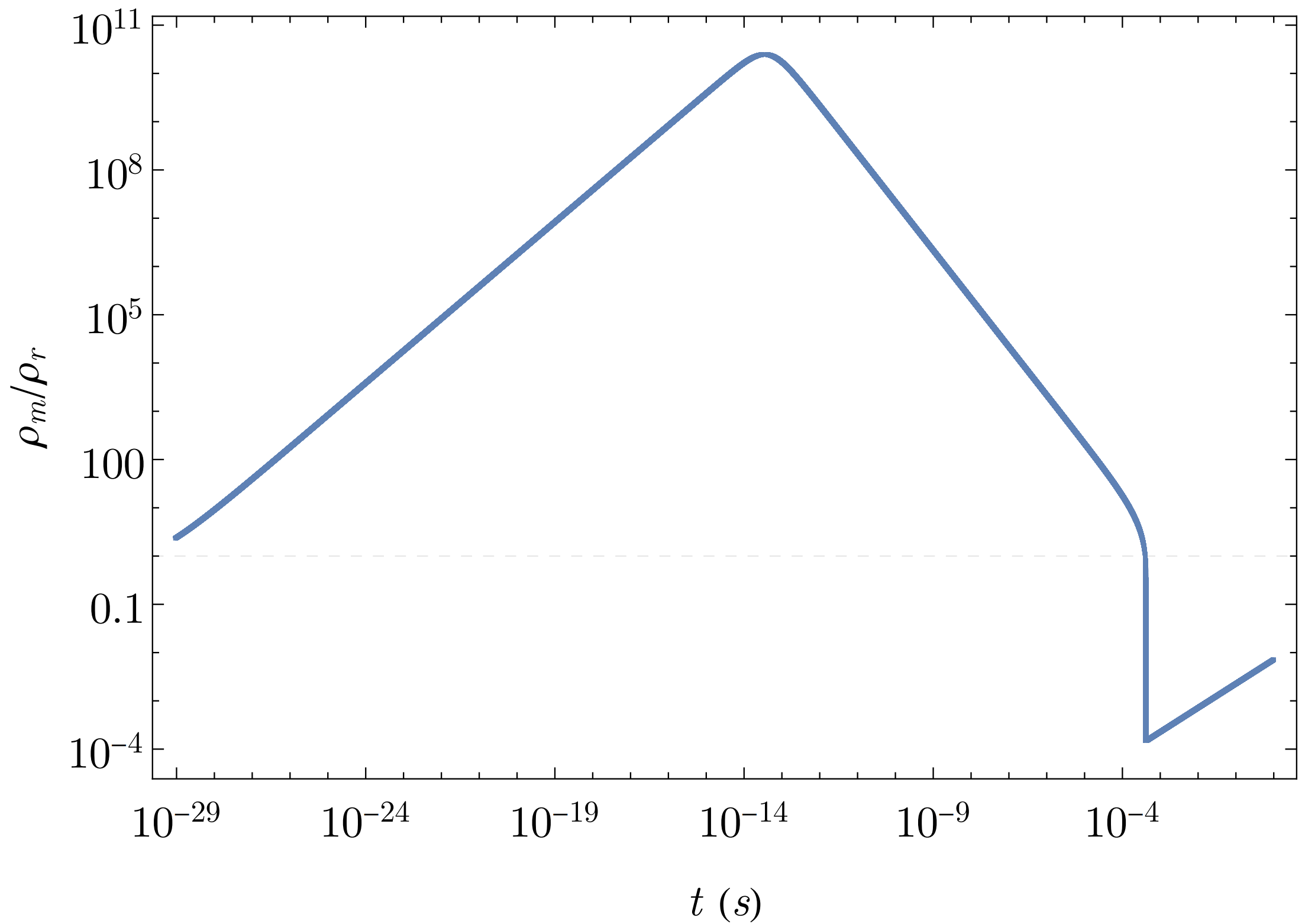}
    \caption{Evolution of $\rho_{\mathrm{m}}/\rho_{\mathrm{r}}$ as a function of time when BHs do not fully evaporate but, instead, form stable relics when approaching the Planck mass. The new regime starts here for $t\approx 2\times 10^{-4}$ s.}
    \label{fig:relics}
\end{figure}

A simple gas of BHs can generate a quite involved cosmological expansion history, as shown in Fig. \ref{fig:relics}. 

%\begin{figure}
%    \centering
%    \includegraphics[width = 0.45 \textwidth]{rhoR relics.pdf}
%    \caption{Caption}
%    \label{fig:enter-label}
%\end{figure}

\subsection{Cosmological constant}

Although the previous discussion would mostly apply to the primordial Universe, it makes sense to add a cosmological constant (assumed to be positive, as measured). This, indeed, remains compatible with the core hypothesis of this work, which consists in studying theoretically the full dynamics in vacuum. The Friedmann equation now reads
\begin{equation}
    H^2=\frac{8\pi G}{3c^2}(\rho_m+\rho_r+\rho_{\Lambda}),
\end{equation}
where $\rho_{\Lambda}\equiv\frac{\Lambda c^2}{8\pi G}$.\\

Should the ratios of the BH and radiation densities be expressed as a function of the scale factor, the addition of a cosmological constant would change nothing at all, whatever its value. As $\rho_m$ and $\rho_r$ are decreasing functions of the scale factor whereas $\rho_\Lambda$ is constant, the latter will inevitably dominate at some point. When using $a$ as the evolution variable, this anyway makes {\it strictly} no difference.\\

\begin{figure}[h]
    \centering
    \includegraphics[width = 0.45 \textwidth]{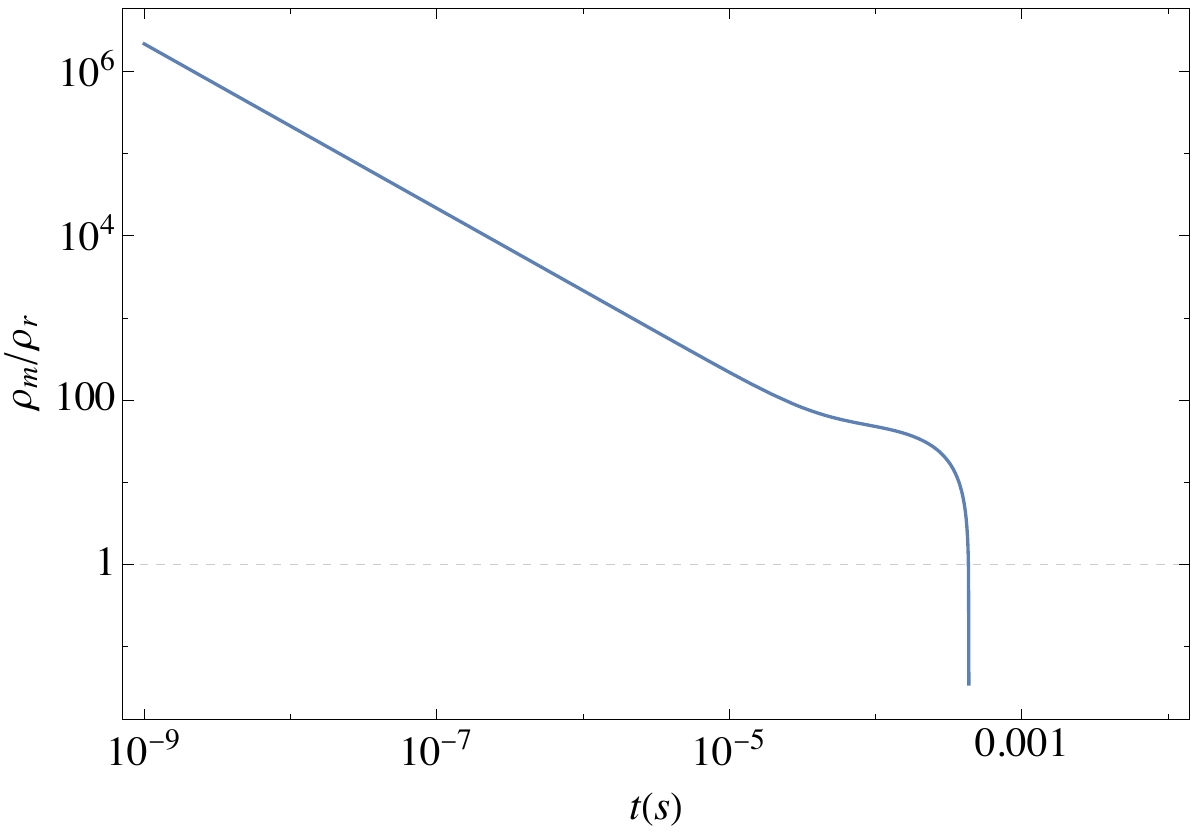}
    \caption{Same ratio $\rho_{\mathrm{m}}/\rho_{\mathrm{r}}$ as on the previous figure but with an additional cosmological constant which begins to dominate at $t=10^{-5}$ s.}
    \label{fig:lambda}
\end{figure}

The situation is obviously different when using the usual cosmic time. At some stage, the cosmological constant dominates. When this happens, the scale factor begins to grow exponentially. As matter is less diluted than radiation, this favors black holes over photons. This is precisely what can be seen in Fig. \ref{fig:lambda} where a vacuum energy density, comparable to the total matter and radiation density at $t=10^{-5}$ s, was added. As expected, when the cosmological constant dominates -- that is for $t\gtrsim  10^{-5}$ s -- the ratio $\rho_m/\rho_r$ decreases more slowly, and eventually increases, until the black holes ``explode". In this final stage, taking place around $t=3\times 10^{-4}$ s for the initial conditions chosen in Fig. \ref{fig:lambda}, the ratio necessarily plunges to zero, whatever the background dynamics.\\

Interestingly, the addition of a cosmological constant can therefore enrich the previously studied evolution. If its value is such that is begins to dominate after the (first) maximum of $\rho_m/\rho_r$ but before the full disappearance of the black holes, it will lead to the second local maximum before the the unavoidable final decrease.\\

The influence of the $\Lambda$-term on the individual densities of BHs and radiation, $\rho_m$ and $\rho_r$, is much less interesting and simply leads to an exponential dilution at the time of dominance of the vacuum energy.

\section{Conclusion}

In this work, we have considered  several situations where Hawking evaporation induces a dynamic competition with classical evolution. Our aim was not to provide ``new physics" results but to highlight some non-trivial conclusions that are not intuitively straightforward.

When considering binary systems of Schwarzschild black holes, we have, in particular, studied exhaustively the evolution of the gravitational wave and graviton frequencies, powers, and cumulative energies for all possible scenarios (inspiraling, outspiraling, and non-monotonic). Most conclusions were not {\it a priori} obvious. We have also established that the full energy transferred to space-time as gravitons is often of the same order and in any case higher than the one transferred as gravitational waves.

When dealing with the expansion of a space filled with a gas of Schwarzschild black holes, we have proven that the matter and radiation energy densities evolve in an intricate way. The main features are universal and do not depend on the initial conditions. Some useful analytical approximations were also provided.

Several developments could be considered in future works. First, it would be interesting to also consider hyperbolic trajectories \cite{Garcia-Bellido:2017qal}. The emission of gravitational waves is very different in this case \cite{Garcia-Bellido:2017knh}. Second, it could make sense to also focus on the graviton emission enhancement taking place if black holes are spinning \cite{Dong:2015yjs}. Finally, possible modifications of the Hawking evaporation happening in the quantum gravity regime (see references in \cite{Barrau:2022eag}) should be taken into account.

\section{Acknowledgements}

The authors would like to thank Cyril Renevey who provided us with the first version of the code initially used for numerical simulations.

\section{Appendix}
Let us review here in details the influence of the non-circularity of the orbits. To this aim, we parameterize the elliptic trajectory by its eccentricity $e$, such that $0 \leq e < 1$, its semi-major axis $a$, and by the polar coordinates $(r, \psi)$ in the plane of the orbit, with the origin located at the position of the center-of-mass. Hence $\psi$ corresponds what is usually called the true anomaly. The polar equation of the ellipse is given by
\begin{equation}
    r = \frac{a(1-e^2)}{1 + e \cos \psi},
\end{equation}
and the second mass moments $M^{ij} \equiv \mu x^{i}(t)x^{j}(t)$ with  $\mu=m_1m_2/(m_1+m_2)$ the reduced mass (in the generic case of a binary systems of objects with masses $m_1$ and $m_2$), and $x^{i}(t)$ the Cartesian coordinates describing the elliptic motion. The are given by
\begin{equation}
    M_{ij}(t) = \mu r^{2} \begin{pmatrix} \cos^{2} \psi & \sin \psi \cos \psi \\ \sin \psi \cos \psi & \sin^{2} \psi \end{pmatrix}.
\end{equation}\\

Restraining ourselves to the case of a binary system made of identical masses, \textit{i.e.} $\mu = m/2$, and introducing the frequency $\Omega_{0}^{2} \equiv \frac{Gm_{\mathrm{tot}}}{a^{3}} = \frac{2Gm}{a^{3}}$, the formulae of \cite{Holgado:2019ndl} become 
\begin{widetext}
\begin{equation}
    \begin{split}
        \dddot{M}_{11} = &\frac{1}{\sqrt{2}} m \Omega_{0}^{3}a^{2} \frac{(1+e\cos \psi)^{2}}{(1-e^2)^{5/2}} \left[ 2 \sin (2\psi) + 3e \sin \psi \cos^{2} \psi \right] - 4 \dot{m} \Omega_{0}^{2} a^{2} (1+e \cos \psi) \cos^{2} \psi \\ &+ \frac{3}{2} \dot{m} \Omega_{0}^{2}a^{2} \frac{(1+e\cos \psi)^{2}}{1-e^{2}} g_{1}(\psi) + \frac{3}{2\sqrt{2}} \ddot{m} \Omega_{0}a^{2} \sqrt{1-e^2} g_{2}(\psi) + \frac{\dddot{m}}{2}a^{2} \left(\frac{1-e^2}{1+e\cos \psi}\right)^{2} \cos^{2} \psi,
    \end{split}
\end{equation}
\begin{equation}
    \begin{split}
        \dddot{M}_{22} = &-\frac{1}{\sqrt{2}} m \Omega_{0}^{3}a^{2} \frac{(1+e\cos \psi)^{2}}{(1-e^2)^{5/2}} \left[ 2 \sin (2\psi) + e \sin \psi (1+3e\cos^{2} \psi) \right] - 4 \dot{m} \Omega_{0}^{2} a^{2} (1+e \cos \psi) \sin^{2} \psi\\ &+ \frac{1}{2} \dot{m} \Omega_{0}^{2}a^{2} \frac{(1+e\cos \psi)^{2}}{1-e^{2}} \tilde{g}_{1}(\psi) + \frac{3}{2\sqrt{2}} \ddot{m} \Omega_{0}a^{2} \sqrt{1-e^2} \tilde{g}_{2}(\psi) + \frac{\dddot{m}}{2}a^{2} \left(\frac{1-e^2}{1+e\cos \psi}\right)^{2} \sin^{2} \psi,
    \end{split}
\end{equation}
\begin{equation}
    \begin{split}
        \dddot{M}_{12} = &\frac{1}{\sqrt{2}} m \Omega_{0}^{3}a^{2} \frac{(1+e\cos \psi)^{2}}{(1-e^2)^{5/2}} \left[ -2 \cos (2\psi) + e \cos \psi (1-3e\cos^{2} \psi) \right] - 2 \dot{m} \Omega_{0}^{2} a^{2} (1+e \cos \psi) \sin (2\psi)\\ &+ \frac{3}{4} \dot{m} \Omega_{0}^{2}a^{2} \frac{(1+e\cos \psi)^{2}}{1-e^{2}} \bar{g}_{1}(\psi) + \frac{3}{2\sqrt{2}} \ddot{m} \Omega_{0}a^{2} \sqrt{1-e^2} \bar{g}_{2}(\psi) + \frac{\dddot{m}}{4}a^{2} \left(\frac{1-e^2}{1+e\cos \psi}\right)^{2} \sin(2\psi),
    \end{split}
\end{equation}
\end{widetext}
where the functions $g_{1}, g_{2}$, etc. are only depending on the true anomaly $\psi$ and on the eccentricity $e$. They are bounded to be smaller that 3. Explicitly, they read, in the approximation that that the precession of the argument of periapsis is  negligible:
\begin{equation}
\begin{split}
    &g_{1} (\psi) \equiv \frac{e^{2} \sin^{2}\psi \cos^{2}\psi}{(1+e\cos \psi)^{2}} - \frac{e \sin \psi \sin(2\psi)}{1 + e\cos \psi} + \sin^{2} \psi, \\
    &\tilde{g}_{1}(\psi) \equiv \frac{e^{2} \sin^{4}\psi}{(1+e\cos \psi)^{2}} + \frac{e \sin \psi \sin(2\psi)}{1 + e\cos \psi} + \cos^{2} \psi, \\
    &\bar{g}_{1}(\psi) \equiv \frac{e^{2} \sin^{2}\psi \sin(2\psi)}{(1+e\cos \psi)^{2}} + \frac{2e \sin \psi \cos(2\psi)}{1 + e\cos \psi} - \sin(2\psi),
\end{split}
\end{equation}
\begin{equation}
\begin{split}    
    &g_{2}(\psi) \equiv \frac{2e \sin \psi \cos^{2}\psi}{1 + e\cos \psi} - \sin(2\psi), \\
    &\tilde{g}_{2}(\psi) \equiv \frac{2e \sin^{3} \psi}{1 + e\cos \psi} - \sin(2\psi), \\
    &\bar{g}_{2}(\psi) \equiv \frac{e \sin \psi \sin(2\psi)}{1 + e\cos \psi} + \cos(2\psi).
\end{split}
\end{equation}
For isotropic mass loss, the periapsis distance $r_{\mathrm{p}} = a(1 - e)$ is constant, so $\dot{r}_{\mathrm{p}}= 0$. When including gravitational radiation reaction, the periapsis distance will indeed decrease, though one can take $\dot{r}_{\mathrm{p}} \approx 0$ on timescales less than the mass-loss timescale $m/ \abs{\dot{m}}$ that is when the gravitational waves coalescence timescale is much longer than the mass-loss timescale. This is the only hypothesis that was made in order to derive these results. It should also be noticed that when enforcing constant masses in to the above formulae, one recovers the canonical expressions of \cite{Peters:1963ux}. If one doesn't make the assumption of \textit{identical} time-varying masses, the corrective terms involve derivatives of the reduced mass $\mu$ and of the total mass $m_\mathrm{tot}$, which may also be thought of as time derivatives of the binary’s chirp mass
$\mathcal{M} \equiv (m_{1}m_{2})^{3/5} m_{\mathrm{tot}}^{-1/5}$.\\

From the inspection of the above formulae, it is clear that the corrective terms are negligible in comparison to the PM canonical expression when \begin{equation}
    \forall \quad 0 < n \leq 3, \qquad \vert m^{(n)}(t)\vert \ll m \Omega_{0}^{n}, 
\end{equation}
with $n$ an integer. This constraint is simply the \textit{slowly-varying mass condition}, which is satisfied during the whole evaporation process except very close to the time of evaporation for which the mass quickly decreases.

\bibliographystyle{apsrev4-1}
\bibliography{article}% Produces the bibliography via BibTeX.

\end{document}